\documentclass[12pt]{article}
\usepackage[utf8]{inputenc}
\usepackage{hyperref}
\usepackage[margin=1in]{geometry}\usepackage{setspace}
\usepackage{color}
\usepackage{braket}
\usepackage{titling}
\usepackage{graphicx}
\usepackage[bf,margin=20pt,font={small}]{caption}
\usepackage{subcaption}
\usepackage{tikz}
\usepackage{youngtab}
\usepackage{epigraph}
\usetikzlibrary{math}
\newcommand{\eps}{\epsilon}
\usepackage{amsmath,amssymb,dsfont}
\allowdisplaybreaks
\newcommand{\tr}{\operatorname{tr}}
\usepackage{tikz}

\definecolor{royalblue}{rgb}{0.00000,0.44700,0.74100}%
\definecolor{royalorange}{rgb}{0.85000,0.32500,0.09800}%
\definecolor{royalyellow}{rgb}{0.92900,0.69400,0.12500}%
\definecolor{purple}{rgb}{0.5804, 0.0, 0.82745098}%
\definecolor{applegreen}{rgb}{0.55, 0.71, 0.0}
\definecolor{bittersweet}{rgb}{1.0, 0.44, 0.37}

\usetikzlibrary{external} 
\tikzsetexternalprefix{images/} 
\tikzset{external/system call={lualatex -shell-escape -halt-on-error -interaction=batchmode -jobname "\image" "\texsource"}}
\newif\ifcompileimages%
\compileimagesfalse%

\DeclareMathAlphabet{\mathpzc}{OT1}{pzc}{m}{it}

\usepackage{putex}

\usepackage{mathtools}
\usepackage{siunitx}

\usepackage{pgfplots}
\pgfplotsset{compat=1.10}
\usepgfplotslibrary{fillbetween}
\usetikzlibrary{patterns}

\institution{UHasselt}{Faculty of Sciences, Hasselt University, Diepenbeek, BE 3590}
\institution{SCGP}{Simons Center for Geometry and Physics, Stony Brook University, Stony Brook, NY 11794}
\institution{NYU}{CCPP, Department of Physics, NYU, New York, NY,  10003, USA}
\institution{MIPT}{Moscow Institute for Physics and Technology, Institute lane 9, Dolgoprudny, Moscow, Russia}

\title{Chaotic RG Flow in Tensor Models}

\authors{Maikel M. Bosschaert\worksat{\UHasselt}, Christian B. Jepsen\worksat{\SCGP}, Fedor K.~Popov\worksat{\NYU,\MIPT
}\footnote{Simons Junior Fellow}
}

\abstract{
We study bi-antisymmetric tensor quantum field theories with $O(N_1)\times O(N_2)$ symmetry. Working in $4-\epsilon$ dimensions we calculate the beta functions up to second order in the coupling constants and analyze in detail the Renormalization Group (RG) flow and its fixed points. We allow $N_1$ and $N_2$ to assume general real values and treat them as bifurcation parameters. In studying the behavior of these models in a non-unitary regime in the space of $N_1$ and $N_2$ we find a point where a zero-Hopf bifurcation occurs. In the vicinity of this point, we provide analytical and numerical evidence for the existence of Shilnikov homoclinic orbits, which induce chaotic behavior in the RG flow of a subset of nearby theories. As a simple warm-up example for the study of chaotic RG flows, we also review the non-hermitian Ising chain and show how for special complex values of the coupling constant, its RG transformations are equivalent to the Bernoulli map.
}

\date{\today}

\begin{document}

\maketitle

\tableofcontents

\section{Introduction}

Science abounds with examples of systems governed by simple rules yet exhibiting marvelously complex behaviors. A broad class of instances of this phenomenon is the occurrence of chaos in dynamical systems. By a dynamical system, we mean a system of autonomous first-order differential equations or discrete maps:
\begin{gather}
\dot{g}_i = \beta_i(g_j)\,, \hspace{20mm} g_i^{(n+1)} = \mathcal{R}_i\left(g_j^{(n)}\right)\,, \label{eq:dynsys}
\end{gather}
where the variables $g_i$ are either real- or complex-valued. The study of chaos in such systems dates back to the work of Henri Poincar\'e\ on the three-body problem. In the Hamiltonian formalism, the trajectories of particles in phase space are described precisely by the kind of first-order differential equations listed in \eqref{eq:dynsys}. In his investigation of the three-body equations of motion, Poincar\'e\ was startled to discover that the solution space is vastly more intricate than he had anticipated, encompassing meandering curves of ever-increasing wiggles and an infinitude of periodic orbits dispersed unevenly in phase space. Since the time when Poincar\'e\ caught his first glimpse of chaos, the characteristic properties by which one can identify chaos have become much better understood. In addition to the presence of an infinite number of periodic orbits with an infinite range of periodicities, sometimes forming complicated fractal structures, chaotic systems are characterized by an extreme sensitivity to initial conditions as well as by the property that open sets of initial states evolve in time to spread out densely in the space of all possible states. For a more detailed discussion of what is meant by the term chaos, we refer the reader to Appendix \ref{appendix:chaos}.

Dynamical systems have a wide range of applications in science and technology, and the emergence of chaos is a commonplace occurrence in these applications, be they planetary orbits, atmospheric convection \cite{lorenz1963deterministic}, string theory \cite{maldacena2016remarks,Gross:2021gsj,PhysRevLett.127.021601,PhysRevB.103.L100302} or population dynamics \cite{arnold2012geometrical}. In modern theoretical physics, an important class of dynamical systems are furnished by the beta functions of quantum field theories (QFTs) and their associated \textit{renormalization group (RG) flows}. Rather than trajectories of particles in phase space, these systems describe the flow of coupling constants in a given theory as we vary the length or energy scale at which we view the theory, but the flow equations remain of the form \eqref{eq:dynsys}.  Consequently, in their most general form, QFTs should admit chaotic RG flows, and already Wilson and Kogut entertained this possibility in their classic review \cite{Wilson:1973jj}. However, the theories typically studied by physicists exclusively exhibit an RG flow of a simpler kind, namely heteroclinic flow between fixed points.  Associated herewith is the idea of universality: we can modify the details of the high-energy (UV) theory and still flow to the same low-energy (IR) theory if we remain in the same basin of attraction. Contrariwise, chaos would spell the doom of universality, with even the tiniest change to the UV theory drastically altering the IR theory. In two, three, and four dimensions it is known that unitarity prevents this kind of behavior by guaranteeing the existence of $c$-, $F$-, or $a$-functions that change monotonically under RG flow \cite{zamolodchikov1986irreversibility,Klebanov:2011gs,Jafferis:2011zi,Casini:2012ei,Komargodski:2011vj,Luty:2012ww}, and the same may be true in higher dimensions. It has further been suggested \cite{Binder:2019zqc} that universality may extend beyond the realm of unitary theories. Nevertheless, the 80s and 90s bore witness to a number of ideas for and examples of chaotic RG flows in certain simple systems \cite{mckay1982spin,berker1984hierarchical,svrakic1982hierarchical,Damgaard:1991zh, Damgaard:1991zb, Dolan:1994wt}, see also \cite{Damgaard:1991zb} for a general discussion. In all these examples, the RG transformations are discrete. The realization of chaos here hinges on the underlying model being non-unitary or involving an unusual hierarchical coupling pattern of spins with no conventional continuum limit, or the chaos arises as an artifact of discrete and approximate RG transformations, in the same manner as the logistic map is chaotic while the solution to the logistic differential equation is monotonic. 

In the paper, we present a family of QFTs of which a subset of non-unitary theories exhibit continuous RG flows that are chaotic. The models arise on analytic continuation of conventional theories by allowing symmetry groups of matrices of non-integer size. The specific models have no concrete experimental motivation and are intended rather to serve as a proof of principle, but non-unitary theories generally, as well as theories with symmetry groups of non-integer size specifically, are capable of describing physical phenomena that can be realized experimentally. The list of non-unitary models of interest in theoretical physics includes such theories as the $q$ state Potts model with $q>4$ \cite{1982RvMP...54..235W,gorbenko2018walking,Gorbenko:2018dtm}, logarithmic CFTs \cite{Gurarie:1993xq,cardy2013logarithmic}, and Liouville theory in dimensions greater than 2 \cite{levy2018liouville}. Furthermore, we observe that analytical continuation of RG flows of conventional field theories provides a method of generating a vast range of dynamical systems. This suggests the possibility of studying systems of interest outside of theoretical physics using QFT methods. As a point in case, we demonstrate in the next section that the Bernoulli map is secretly identical to the RG transformations of the 1d Ising model at special complex values of the coupling constant.

In general, it is very difficult to conclusively prove the presence of chaos in a system, but some tools are available. One method is to map a system onto one of the few well-studied systems that are known to be chaotic. We apply this method in section \ref{sec:Ising}, where, as a toy model with chaotic behavior, we analyze the complexified Ising chain, which was previously studied and shown to be chaotic in \cite{Dolan:1994wt}. For continuous dynamics, a set of necessary conditions for the onset of chaotic dynamics, involving the presence of a homoclinic orbit, was put forward by Shilnikov \cite{shilnikov1965case}. We review his construction in section \ref{sec:Shilnikov}. Subsequent to Shilnikov's discovery, mathematicians were able to show that his conditions are met generically in the vicinity of certain kinds of bifurcations\cite{baldoma2020hopf}. In section \ref{sec:QFT}, we present a tensor model with $O(N_1)\times O(N_2)$ symmetry whose beta functions undergo such a bifurcation at the special values $N_1=2.521$, $N_2=1.972$, and we provide numerical evidence that there exists a Shilnikov homoclinic orbit among the RG trajectories of the model, thereby establishing that this QFT exhibits chaotic RG flow.

\section{The Complex Ising Chain and the Bernoulli Map}
\label{sec:Ising}
In this section, as an illustration of the ideas of chaos discussed in the introduction and as an example of how universality extends beyond unitarity but breaks down in special chaotic regions, we will consider the one-dimensional Ising model with a complex coupling. This model and its chaotic behavior was also studied in \cite{Dolan:1994wt}. Once coupling constants in a theory are complex, all sorts of RG flows become possible. For instance, it is quite easy to find limit cycles \cite{faedo2021multiple}. The consideration of complex RG fixed points was previously proposed in \cite{Gorbenko:2018dtm,Gorbenko:2018ncu} and further carried out in the context of tensor models in \cite{Giombi:2017dtl}.  The complex Ising model in this section serves as a warm-up for the more intricate model in section \ref{sec:QFT}, where we will realize chaos in the RG flow of real-valued coupling constants.

In the absence of an external magnetic field, the Hamiltonian of the one-dimensional Ising model is given by
\begin{gather}
  H = g \sum_i \sigma_i \sigma_{i+1}\,,
\end{gather}
where $i$ runs over some set of integer labels, and the spin variables $\{\sigma_i\}$ can assume the values $\pm 1$. The  1$d$ Ising model has been studied extensively in the literature, and there are numerous methods to solve it numerically and analytically. The system does not exhibit a phase transition, and we can compute the correlation functions exactly. The most common method of solving the model involves the use of a transfer matrix, which can also be used to implement the RG flow of the model. The idea is the following: the partition function is given by
\begin{gather}
 Z = \sum_{ \left\{ \sigma_i = \pm 1\right\}} \prod_i C e^{-g \sigma_i \sigma_{i+1}}\,,
 \label{Z}
\end{gather}
where we allow for a normalization constant $C$. Supposing we have $N$ spin variables $\sigma_i$ and impose periodic boundary conditions $\sigma_i=\sigma_{N+i}\,$, we notice that we can introduce a matrix $T$ such that
\begin{gather}
T = C\begin{pmatrix}
 e^{-g} & e^{g}\\
 e^{g} & e^{-g}
\end{pmatrix},\hspace{20mm} Z  = \operatorname{tr}\left[ T^N\right] \,. \label{eq:partfunc}
\end{gather}
Now let us carry out one step of an RG flow by integrating out some degrees of freedom of the model, for example the spins with odd indexes in the chain. It is possible to express the partition function in terms of the remaining degrees of freedom by an equation still of the form \eqref{Z}, except that the coupling constant $g$ has to be replaced by a different coupling constant $g_1$, and $C$ by a new normalization constant $C_1$: 
\begin{gather}
 Z = \sum_{ \left\{ \sigma_i = \pm 1\right\}} \prod_{i \text{ even}} C_1e^{-g_1 \sigma_i \sigma_{i+1}}\,.
 \label{Z1}
\end{gather}
The new constants $g_1$ and $C_1$ can be computed by noticing that integrating out the odd spins formally leads to the replacement $T\to T^2 \equiv T_1$. Demanding that $T_1$ has the same form as the original transfer matrix $T$ gives the equation:
\begin{gather}
C^2\begin{pmatrix}
  e^{-2g}+e^{2g} & 2\\
  2 & e^{-2g}+e^{2g}
\end{pmatrix}
=
C_1\begin{pmatrix}
 e^{-g_1} & e^{g_1}\\
 e^{g_1} & e^{-g_1}
\end{pmatrix}
\,.
\end{gather}
This equation implies that the RG step relates the old coupling $g$ to the new coupling $g_1$ via the relation
\begin{gather}
  e^{-2g_1} =\frac{(T^2)_{11}}{(T^2)_{12}}=\frac12\left(e^{-2g}+e^{2g} \right) = \cosh 2g\,.
\end{gather}
For convenience we introduce a new parameter $z=e^{-2g}$, in terms of which the RG step assumes the simple form
\begin{gather}
 z_1 = \frac12\left(z + \frac{1}{z}\right)  \equiv R(z)\,.
\end{gather}
Let us first consider the unitary Ising model, which means we take $z$ to be real and positive. In this case, the sequence
$z_n = R(z_{n-1})$
is convergent. To see this, note that for any positive real, $z$ we have that $z_1 = \frac12\left(z+\frac{1}{z}\right) \geq 1$.
Furthermore, for any $z_n>1$, we have that $z_{n+1} = R(z_n) \leq \frac{1}{2}\left(z_n+1\right) < z_n$. Hence, the sequence $z_1$,$z_2$,$z_3$,$\ldots$ is decreasing, and since it is also bounded below, it is convergent. 
The fixed point that the sequence converges to is situated at $z=1$, i.e. $g=0$, corresponding to a high temperature fixed point.

Suppose now that we do not constrain ourselves to unitary theories and allow $g$ and $z$ to assume complex values. By an argument analogous to the above, one can show that as long as $\operatorname{Re} z \neq 0$, the sequence remains convergent, converging to the value $\operatorname{sign}(z)$. However, as we will now demonstrate, when $z$ is imaginary, the behavior of the sequence changes drastically and chaos emerges. It is not hard to see that when $z_n\equiv i x_n$ is imaginary, then $\widetilde{R}(i x_n)\equiv ix_{n+1}$ is also imaginary, and we have
\begin{gather}
x_{n+1} = \tilde{R}(x_{n})\,,\hspace{20mm} \widetilde{R}(x) \equiv \frac12\left(x-\frac{1}{x}\right)\,.
\end{gather}
We now introduce a parameter $t_n \in \left[0,1\right)$ related to $x_n$ via the equation $x_n = \tan \big(\pi(t_n-\frac12)\big)$. Then the RG step acts on $t_n$ as
\begin{gather}
t_{n+1} = 2 t_n \mod 1\,. \label{tRG}
\end{gather}
This map is known as the dyadic map or the Bernoulli map and was introduced 
by R\'enyi \cite{renyi1957representations} as part of a larger class of transformations that he proved to be ergodic. Despite the simplicity of the map, it exhibits all the characteristics of chaotic flow. One immediate observation is that the map has a non-zero Lyapunov exponent: $\delta t_{n+k} = 2^k \delta t_n$ for sufficiently small $\delta t_n$, although this behavior breaks down at large values of $k$ since $t$ is constrained to a finite interval. It is also not hard to see that for any finite interval $I \subset [0,1]$ of initial values of $t_0$, the image of $I$ under repeated RG steps will eventually spread out over the whole interval $[0,1]$. Furthermore, suppose we decompose the starting value $t_0$ of $t$ in a binary expansion:
\begin{gather}
t_0=\sum_{i=1}^\infty a_i\,2^{-i}\,, 
\end{gather}
where $a_i \in \{0,1\}$. The semi-infinite sequence $A=\left\{a_i\right\}^\infty_{i=1}$ provides all information concerning the initial state and subsequent evolution of the system. And the application of an RG step \eqref{tRG} corresponds to discarding $a_1$ and shifting $a_{i+1} \to a_{i}$. From this point of view, we can exactly predict the evolution of the system when the initial state is known exactly, but any variation whatsoever of the initial state will eventually lead to the largest possible fluctuations in future values of the state of the system. Thus, if $t_0 \in \mathbb{Q}$, the sequence $A$ will be periodic after a finite number of initial digits, meaning that the RG flow becomes cyclic. More precisely, we can express any $t_0 \in \mathbb{Q}$ as a reduced fraction $t_0 = 2^m\frac{q}{r}$ where $m,q\in \mathbb{Z}$, $r\in\mathbb{N}$, and gcd$(q,r)=$gcd$(q,2)=$gcd$(r,2)=1$, in which case the periodicity of the RG flow is simply $r$, while max$(-m,0)$ equals the number of initial digits in $A$ before the sequence becomes periodic. From the above it follows that all periodicities are realized for some initial values of $t$ and that any finite interval of initial values induces RG flows with an infinite set of different periods. Meanwhile, if $t_0 \notin \mathbb{Q}$, then the sequence $A$ does not have a limit cycle. Moreover, one can show that the set $\mathfrak{t}=\left\{t_n\right\}^\infty_{i=1}$ will be dense in the interval $I$.  Since both the rational and irrational numbers are dense in the space of values for $t_0$, it requires infinite precision to determine whether a given initial state gives rise to periodic flow or not. 

\begin{figure}
    \centering
  \includegraphics[width=0.68\textwidth]{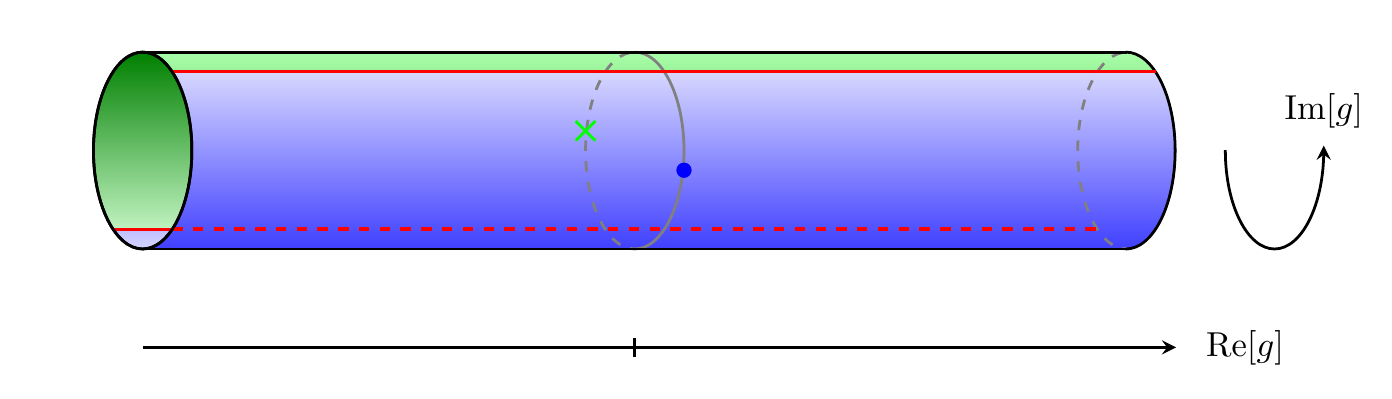}
    \caption{Phase diagram of the complexified Ising chain. The blue region indicates the basin of attraction of the trivial fixed point at $g=0$, while the green region indicates the basin of attraction of the imaginary fixed point at $g=\frac{i\pi}{2}$. The two basins are separated by a chaotic region drawn in red.}
    \label{fig:Ising}
\end{figure}

The full phase diagram of the complexified Ising chain is shown in Figure \ref{fig:Ising}. Since $g$ appears in an exponent accompanied by a factor of plus or minus one, we can restrain the imaginary values of $g$ to the interval $(-\frac{i\pi}{2},\frac{i\pi}{2}]$ so that $g$ is valued over an infinite complex cylinder. One way to understand the phase diagram is to think in terms of the eigenvalues $\lambda_+$ and $\lambda_-$ of the transfer matrix $T$:\footnote{We would like to thank Vladimir Rosenhaus for pointing out this interpretation}
\begin{align}
\lambda_+ = 2C \cosh(g)\,,
\hspace{20mm}
\lambda_- = -2C \sinh(g)\,.
\end{align}
The partition function, being the trace of the $N^{\rm th}$ power of the transfer matrix, can be expressed as
\begin{align}
Z=\lambda_+^N+\lambda_-^N\,.
\end{align}
The blue region in Figure \ref{fig:Ising} corresponds to the range $-\frac{\pi}{2}< \Im[g] <\frac{\pi}{2}$, where $|\lambda_+|>|\lambda_-|$. This means that for large systems, $\lambda_+$ dominates the free energy: 
\begin{align}
F=\log Z \approx N \log \big[C\cosh (g)\big]\,,
\end{align} 
which is the regular behavior of the Ising model. Meanwhile, the green region is the regime where $|\lambda_-|>|\lambda_+|$ so that $\lambda_-$ dominates the free energy. When $g=x+i\frac{\pi}{4}$ for some real number $x$, then $|\lambda_1|=|\lambda_2|$, and we are in the chaotic regime. In this case, for a modulus and phase given by
\begin{align}
\rho = \sqrt{\frac{\cosh(2x)}{2}}\,,\hspace{20mm} \phi = \arctan\Big[\tanh(x)\Big]\,,
\end{align}
we have that the eigenvalues and free energy equal
\begin{align}
\lambda_{\pm}=\rho e^{\pm i\phi}\,, \hspace{20mm} F = N\log(\rho)+\log\Big[2\cos(N\phi)\Big].
\end{align}
Because of the phase in the argument of the cosine, the second contribution could be arbitrarily large, depending on the exact number of sites in the Ising model. Moreover, let us consider the two-point functions and its large $N$ behavior when $\left|\lambda_+\right| > \left|\lambda_-\right|$:
\begin{gather}
\braket{\sigma_i \sigma_{i+k}} = \frac{\tr\left[T^{N-k}\sigma_3 T^k \sigma_3 \right]}{\tr\left[T^N \right]} = \frac{\lambda_+^k \lambda_-^{N-k}+\lambda_-^k \lambda_+^{N-k}}{\lambda_+^N+\lambda_-^N}  \approx \left(\frac{\lambda_{-}}{\lambda_{+}}\right)^k, \quad N\gg k\,.
\end{gather}
This correlation function behaves smoothly in the thermodynamic limit $N\to \infty$.  Meanwhile, on the special line $g= x + i \frac{\pi}{4}$,
\begin{gather}
\braket{\sigma_i \sigma_{i+k}} = \frac{\cos\, (N-2 k ) \phi}{\cos N \phi},
\end{gather}
which is highly sensitive to the total number of the sites and does not admit a simple thermodynamic limit. 

Having closely studied this simple chaotic chain, it is not hard to conceive the implications of  chaotic RG transformation in systems in higher dimensions or composed of different types of spin-sites. Roughly speaking, while an ideal gas is well-described simply be specifying temperature, pressure, and volume, for a {\it RG chaotic} gas composed of a macroscopic number of particles it would require knowledge of on the order of $10^{23}$ parameters to make accurate predictions.

\section{Baker's map, the Smale Horseshoe, and Shilnikov Homoclinic Orbits}
\label{sec:Shilnikov}
\begin{figure}
    \centering
  \includegraphics{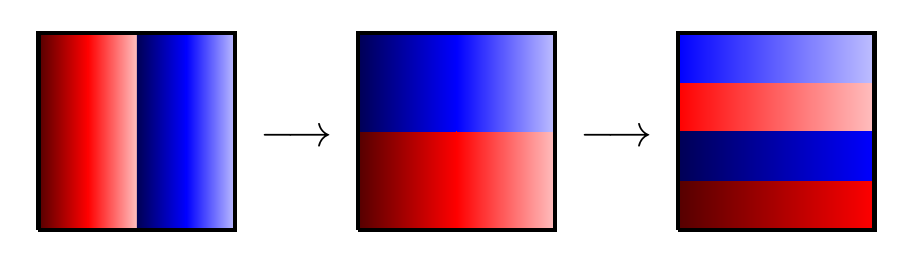}
    \caption{Baker's Map}
    \label{fig:baker}
\end{figure}

In this section, we review one of the few general tools for diagnosing chaos in a continuous dynamical system: Shilnikov homoclinic orbits. In order to properly understand these orbits, we will need to review certain facts concerning chaos in discrete dynamical systems. Consider therefore the discrete map known as the baker's map \cite{hopf1937erg}, which acts on the unit square $I = \left\{(x,y): 0 \leq x,y \leq 1 \right\}$ as
\begin{gather}
B(x, y) =\left(2x - \left\lfloor 2x \right\rfloor,\ \frac{ y + \left\lfloor 2x\right\rfloor }{2}\right),\quad B:I \to I\,.
\end{gather}
On an intuitive level, we simply stretch the unit square by a factor of two in the $x$ direction and squeeze it by a factor of two in the $y$ direction such that the total area is preserved, and afterwards we cut off the right half of the stretched shape and place it on top of the left half\footnote{The name baker's map derives from the similarity of this process with the kneading of dough.}. See Figure \ref{fig:baker}. 
It can be seen that the map is chaotic by associating to each point $(x,y)$ of the unit square $I$ an infinite sequence of numbers $\sigma =\left\{ \sigma_i \right\}^\infty_{i=-\infty}$, $\sigma_i \in \{0,1\}$, defined via the binary expansions of $x$ and $y$ as
\begin{gather}
x = \sum\limits^\infty_{i=1} \frac{\sigma_i}{2^{i}}\,,
\hspace{20mm}
y= \sum\limits_{i=0}^{\infty} \frac{\sigma_{-i}}{2^{i+1}}\,.
\end{gather}
The baker's map acts on $\sigma$ by shifting each number one step to the right:
\begin{gather}
\tilde{\sigma} \equiv   B(\sigma)  = \left\{ \tilde{\sigma}_i=\sigma_{i-1} \right\}^\infty_{i=-\infty}\,.
\end{gather}
From this fact one can immediately infer certain chaotic properties of the system as we did for the Bernoulli map in the previous section. For instance, given any sequence $\sigma_T$ that is periodic with a given period $T$, the orbit of $\sigma_T$ under repeated applications of the map $B$ will in turn be periodic with period $T$. And the set of all points $(x,y)$ with a periodic sequence $\sigma$ is dense in the unit square. If we take $x$ and $y$ to be irrational, the orbit of $(x,y)$ never returns to the original point. And the set of points $(x,y)$ with irrational $x$ and $y$ is also dense in the unit square. Hence, the fate of any orbit of the baker's map is infinitely sensitive to initial conditions.
\begin{figure}
    \centering
  \includegraphics[width=0.7 \textwidth]{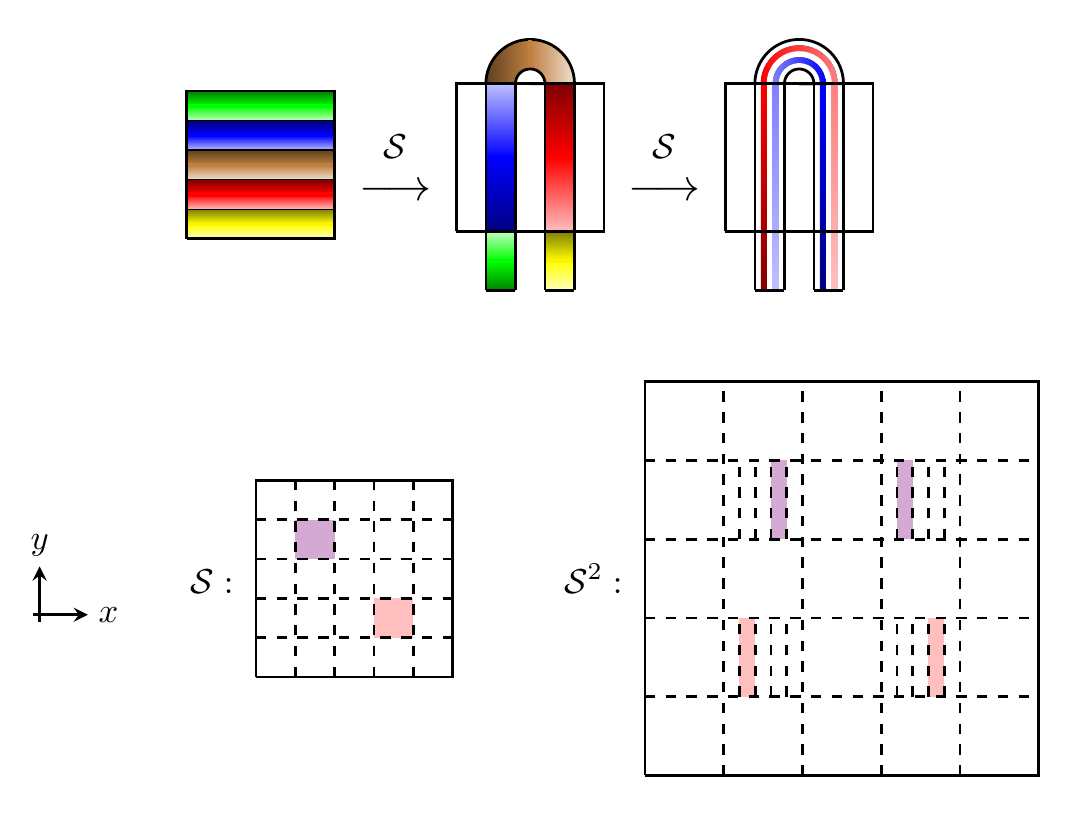}
    
    \caption{Above: the Smale Horseshoe map. We extend the square in the vertical direction and then bend it in the middle. Below: Regions that necessarily contain fixed points under one and two applications of the horseshoe map. The total point set that is periodic under the horseshoe map forms an infinite fractal set.}
    \label{fig:SHmap}
\end{figure}

The baker's map is a limiting case of a more general chaotic map known as the Smale horseshoe map \cite{smale1967differentiable} and depicted in Figure \ref{fig:SHmap}. We will denote this map by $\mathcal{S}$. One can argue that the map is chaotic by showing that it contains a fractal set of periodic orbits. To see why this is so, consider the two highlighted regions on the lower left in Figure \ref{fig:SHmap}. Each of these regions must necessarily contain a fixed point under the horseshoe map. Take for example the \textcolor{pink}{pink} region. In this region, red is mapped to red in the color scheme of the top part of Figure \ref{fig:SHmap}. Since the pre-image in this region sweeps through the entire red palette, there must necessarily be some horizontal line where image and preimage of $\mathcal{S}$ have identical hues, i.e. where the $y$-component is unchanged by $\mathcal{S}$. By instead drawing the horseshoe map with a color palette that runs from left to right, one can similarly argue that there is exists a vertical line in the \textcolor{pink}{pink} region, where the $x$-component is unchanged by $\mathcal{S}$. At the intersection of the vertical and horizontal lines just described, there must be a fixed point. By the same argument, it is not hard to see that $\mathcal{S}^2$ must have a fixed point in each of the four regions highlighted in the lower right of Figure \ref{fig:SHmap}, and that in general $\mathcal{S}^n$ must have $2^n$ fixed points. From this line of reasoning, it becomes apparent that $\mathcal{S}$ contains a Cantor set of periodic orbits with all possible periods, which furnishes evidence to the fact that the Smale horseshoe map is chaotic. Furthermore, it can be rigorously proved that any map topologically equivalent to the Smale horseshoe map is also chaotic \cite{guckenheimer2013nonlinear}.

So far we have reviewed discrete dynamical systems, but we now turn to continuous dynamical systems described by a set of autonomous first-order differential equations. For such systems, it is also possible to establish the presence of chaos via the horseshoe map. To do this, one needs a way of deriving a two-dimensional discrete map from a system of differential equations. This is achieved with the Poincar\'e map: given a continuous dynamical system in more than two dimensions, we consider some fixed two-dimensional surface $S$ that is traversed by the trajectories of the dynamical system. For any point $s_0 \in S$ there is a unique solution curve that passes through $s_0$. Suppose this curve is such that we can follow it forward in time starting at $s_0$ until we find that it crosses $S$ again at some point $s_1$. For all such points $s_0$, the Poincar\'e map $\pi$ is defined by the relation $\pi(s_0) = s_1$.

\begin{figure}
    \centering
  \includegraphics{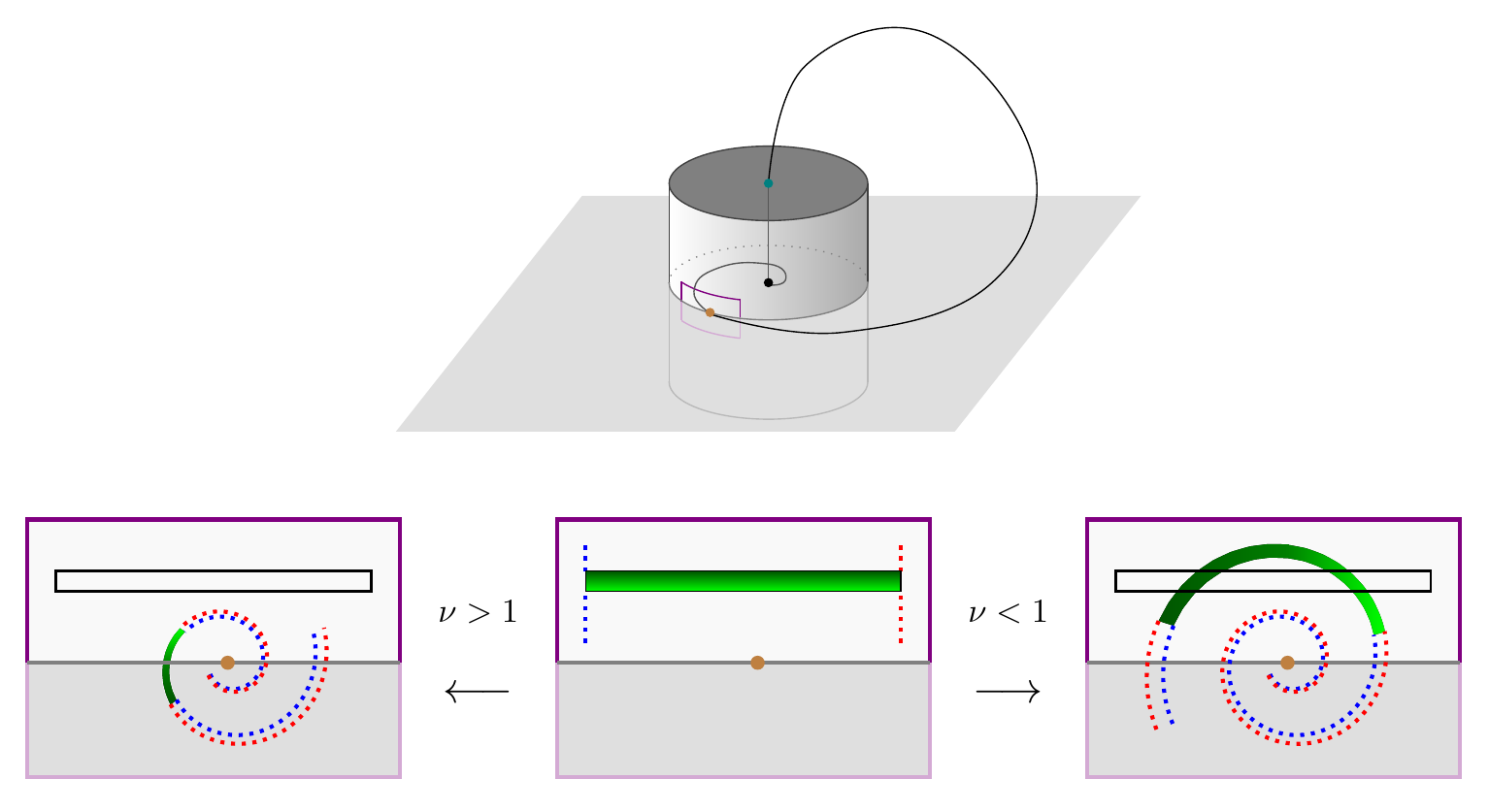}
    \caption{Homoclinic orbit and the Poincar\'e map \eqref{eq:ShilProof} that it induces for different values of the ratio $\nu = \frac{\rho}{\gamma}$. For $\nu<1$ it is always possible to find a subregion that undergoes a Smale horseshoe map as shown on the right. For $\nu>1$ this is not the case.}
    \label{fig:Shilnikov}
\end{figure}

From the above discussion, we see that whenever in a continuous dynamical system we are able to find a Poincar\'e map that induces a Smale horseshoe map, then we know that the system is chaotic. A simple set of sufficient criteria for making this determination was found by Shilnikov \cite{shilnikov1965case}, who managed to prove that a three-dimensional system is chaotic if it contains a homoclinic orbit originating and terminating at a fixed point such that the stability matrix at this point has a pair of complex conjugate eigenvalues $\lambda_\pm = -\rho \pm i \omega$ and a real eigenvalue $\lambda_3=\gamma > \rho$. The proof runs roughly as follows: In the vicinity of the fixed point, the dynamical system can be described by coordinates $x$, $y$, and $z$ satisfying the equations
\begin{gather}
    \begin{array}{rclcl}
\dot x &=&  -\rho x - \omega y &+& F_1(x,y,z) \,,\\
\dot y &=& ~~\omega x - \rho y  &+& F_2(x,y,z)\,,\\
\dot z &=& \gamma z &+& F_3(x,y,z)\,,
\end{array}
\end{gather}
where $F_i(x,y,z)$, $i \in \{1,2,3 \}$, are functions of second order or higher in the coordinates.
Suppose now we surround the fixed point by a cylindrical shell whose curved surface is given by $C=\left\{(x,y,z): x^2+y^2 = r^2 , |z| \leq h \right\}$, see Figure \ref{fig:Shilnikov}. The homoclinic orbit enters the cylinder through the curved surface at some point $p$ (marked in \textcolor{brown}{brown} in Figure \ref{fig:Shilnikov}) and exits at some point $q$ (marked in \textcolor{teal}{teal} in Figure \ref{fig:Shilnikov}) on the flat top of the cylinder. By taking $r$ and $h$ to be very small, we can assume that $q$ is situated on the $z$-axis and that $p$ lies in the $(x,y)$-plane so that we can choose to write $q=(0,0,h)$ and $p=(r,0,0)$. Disregarding the higher-order terms $F_i(x,y,z)$, we determine that the solution curves near $p$ are given in terms of local coordinates $(\zeta,\theta)$ by
\begin{align}
    x(t) =&\, r e^{-\rho t} \cos(\omega t + \theta)\,, \notag  \\
    y(t) =&\, r e^{-\rho t} \sin(\omega t +\theta)\,,\\
    z(t) =&\, \zeta e^{\gamma t}\,. \notag 
\end{align}
In this approximation, any given solution curve, intersecting $C$ at time $t=0$ at a positive value $\zeta$ of $z$, subsequently intersects the plane $z=h$ at the coordinates
\begin{gather}
\left\{
\begin{matrix}
    x(\theta,\zeta) = r\left(\frac{\zeta}{h}\right)^{\frac{\rho}{\gamma}} \cos\left(\theta - \frac{1}{\lambda} \log \frac{\zeta}{h}\right),\\
    y(\theta,\zeta) = r\left(\frac{\zeta}{h}\right)^{\frac{\rho}{\gamma}} \sin\left(\theta - \frac{1}{\lambda} \log \frac{\zeta}{h}\right).
\end{matrix}\right.    
\label{eq:fPM}
\end{gather}
For tiny values of $x$ and $y$, the solution curve will be close to the homoclinic orbit and will flow back to again intersect the surface $C$ at some point close to the point $p$. Thereby, the flow induces a Poincar\'e map that maps points on $C$ with $z > 0$ back to $C$. To first order in $x$ and $y$ we can approximate this map by a linear transformation:
\begin{gather}
    \left\{
    \begin{matrix}
    \zeta' =A x(\theta,\zeta) + B y(\theta,\zeta),\\
    \theta' = C x(\theta,\zeta) + D y(\theta,\zeta).
    \end{matrix} \right. \label{eq:sPm}
\end{gather}
Combining the equations \eqref{eq:fPM} and \eqref{eq:sPm} we arrive at the following map:
\begin{gather}
        \left\{
    \begin{matrix}
    \zeta' = r\left(\frac{\zeta}{h}\right)^{\frac{\rho}{\gamma}} \left[ A  \cos\left(\theta - \frac{1}{\lambda} \log \frac{\zeta}{h}\right) + B  \sin\left(\theta - \frac{1}{\lambda} \log \frac{\zeta}{h}\right)\right],\\
    \theta' = r\left(\frac{\zeta}{h}\right)^{\frac{\rho}{\gamma}}\left[ C  \cos\left(\theta - \frac{1}{\lambda} \log \frac{\zeta}{h}\right) + D  \sin\left(\theta - \frac{1}{\lambda} \log \frac{\zeta}{h}\right)\right].
    \end{matrix} \right. \label{eq:ShilProof}
\end{gather}
For generic values of $A$, $B$, $C$, and $D$, vertical lines of constant $\theta$ and increasing $\zeta$ are mapped to spirals that wind outward around the origin as sketched in figure  \ref{fig:Shilnikov}. Now, the key observation is that for $\nu \equiv \frac{\rho}{\gamma} < 1$, when $\frac{\zeta}{h}$ is small as required for the approximation \eqref{eq:ShilProof} to be valid, points $(\zeta,\theta)$ are mapped to points $(\zeta',\theta')$ that are farther from the origin. For this reason, drawing a plot of the map \eqref{eq:ShilProof} as in Figure \ref{fig:Shilnikov}, one finds that when the Shilnikov condition
\begin{align}
\nu < 1 \label{eq:ShilnikovCondition}
\end{align}
is satisfied, then generically it is always possible to find a region of $C$ that undergoes a Smale horseshoe map. This concludes our proof sketch of Shilnikov's theorem. In the next section, we present examples of Shilnikov homoclinic orbits that occur in the context of RG flow.

\section{Chaotic Bi-Antisymmetric Tensor Model}\label{sec:model}
\label{sec:QFT}
In this section, we present a family of tensor models \cite{Klebanov:2016xxf,Klebanov:2018fzb} with $O(N_1)\times O(N_2)$ symmetry and show that for special non-integer values of $N_1$ and $N_2$, the RG flows of the models become chaotic.\footnote{ For an interesting large $N$ limit of a model with $SO(N_1)\times SO(N_2)$ we refer the reader to \cite{Chaudhuri:2020xxb,kapoor2021bifundamental}} In studying the RG flow of a model whose symmetry group is of non-integer size, we follow the program of \cite{Jepsen:2020czw} and \cite{Jepsen:2021rhs}, which in this setting discovered the existence of RG limit cycles and homoclinic orbits. Generally, models with non-integer dimensional symmetry groups can be studied numerically and analytically \cite{Binder:2019zqc,gorbenko2018walking}, can describe real physical phenomena \cite{de1979scaling}, and avoid the famous $a,c,F-$ theorems \cite{zamolodchikov1986irreversibility,Komargodski:2011vj,Klebanov:2011gs,Jafferis:2011zi}  at the price of non-unitarity. For to study non-monotonic RG-flow it is necessary to consider regimes where the Zamolodchikov metric has negative eigenvalues, which implies that  some of the two-point functions are equipped with negative coefficients.

The main advantage to treating $N_1$ and $N_2$ as real numbers that we can tune to any desired value lies in the fact that this approach allows us to apply theorems of bifurcation theory, which provide one of the sole means of firmly establishing that a dynamical system is chaotic. The idea to treat $N$ in the $O(N)$ model as a bifurcation parameter was suggested already 
in Ref. \cite{Damgaard:1991zb}. In 
Ref. \cite{ibanez1995sil} proved that in a general class of 4-parameter dynamical systems, it is possible to tune the parameters to such values that generically Shilnikov chaos is guaranteed to occur. 
In Ref. \cite{ibanez2005shil} these results were improved, by proving that Shilnikov homoclinic orbits arise in the vicinity of a certain kind of codimension-3 bifurcation. Finally, in 2020 \cite{baldoma2020hopf} provided rigorous evidence for the generic presence of Shilnikov chaos in dynamical systems that undergo a subtype of the codimension-2 bifurcation known as the zero-Hopf (ZH) bifurcation. In the tensor model we shall presently shortly, this kind of bifurcation is realized.

We consider a family of scalar models described by rank-four tensor fields $\phi^{ab}_{\alpha\beta}$ where the upper indices run from one to $N_1$: $a_1,a_2 \in\{1,...,N_1\}$ and belong to the antisymmetric representation of the $O(N_1)$ group, while the lower indices run from one to $N_2$: $\alpha,\beta\in\{1,...,N_2\}$ and also belong to the antisymmetric representation of the $O(N_2)$ group:
\begin{align}
\phi^{ab}_{\alpha\beta} = -\phi^{ba}_{\alpha\beta} = \phi^{ba}_{\beta\alpha}\,.
\end{align}
We work in $d=4-\epsilon$ Euclidean dimensions, where quartic interactions in the fields are marginally relevant. Unlike the case for $N_1$ and $N_2$, the number of spacetime dimensions $d$ being non-integer valued is not crucial to the realization of RG chaos in this section. The effect of working in slightly less than four dimensions is to introduce small linear terms into the beta functions for the coupling constants we introduce below. The same effect can be achieved by coupling the model to gauge fields, but for simplicity of presentation we do not adopt this approach but work in $4-\epsilon$ dimensions instead.

In total the family of models has eight marginal, quartic interactions $O_i$, and we take the action to be
\begin{align}
S = \int d^dx\bigg[\frac{1}{2}(\partial_\mu \phi^{ab}_{\alpha\beta})^2
+\frac{1}{4!}\sum_{i=1}^8g_i\,O_i(x)\bigg]\,, \label{eq:action}
\end{align}
where the operators $O_i$ are given by
\begin{align}
&O_1= \Big(\phi^{ab}_{\alpha\beta}\,\phi^{ab}_{\alpha\beta}\Big)^2\,,
\hspace{28mm}
O_2= \phi^{ab}_{\alpha\beta}\,\phi^{ab}_{\gamma\delta}\,\phi^{cd}_{\gamma\delta}\,\phi^{cd}_{\alpha\beta}\,,
\notag
\\
\label{eq:operators}
&O_3= \phi^{ab}_{\alpha\beta}\,\phi^{ab}_{\gamma\beta}\,\phi^{cd}_{\gamma\delta}\,\phi^{cd}_{\alpha\delta}\,,
\hspace{22mm}
O_4= \phi^{ab}_{\alpha\beta}\,\phi^{cb}_{\alpha\beta}\,\phi^{cd}_{\gamma\delta}\,\phi^{ad}_{\gamma\delta}\,,
\\
&O_5= \phi^{ab}_{\alpha\beta}\,\phi^{ab}_{\gamma\delta}\,\phi^{cd}_{\alpha\delta}\,\phi^{cd}_{\gamma\beta}\,,
\hspace{22mm}
O_6= \phi^{ab}_{\alpha\beta}\,\phi^{cd}_{\alpha\beta}\,\phi^{ad}_{\gamma\delta}\,\phi^{cb}_{\gamma\delta}\,,
\notag
\\
&O_7= \phi^{ab}_{\alpha\beta}\,\phi^{cb}_{\gamma\beta}\,\phi^{cd}_{\gamma\delta}\,\phi^{ad}_{\alpha\delta}\,,
\hspace{22mm}
O_8= \phi^{ab}_{\alpha\beta}\,\phi^{cb}_{\gamma\delta}\,\phi^{cd}_{\gamma\beta}\,\phi^{ad}_{\alpha\delta}\,.
\notag
\label{eq:oper}
\end{align}
\begin{figure}
\centering
  \includegraphics{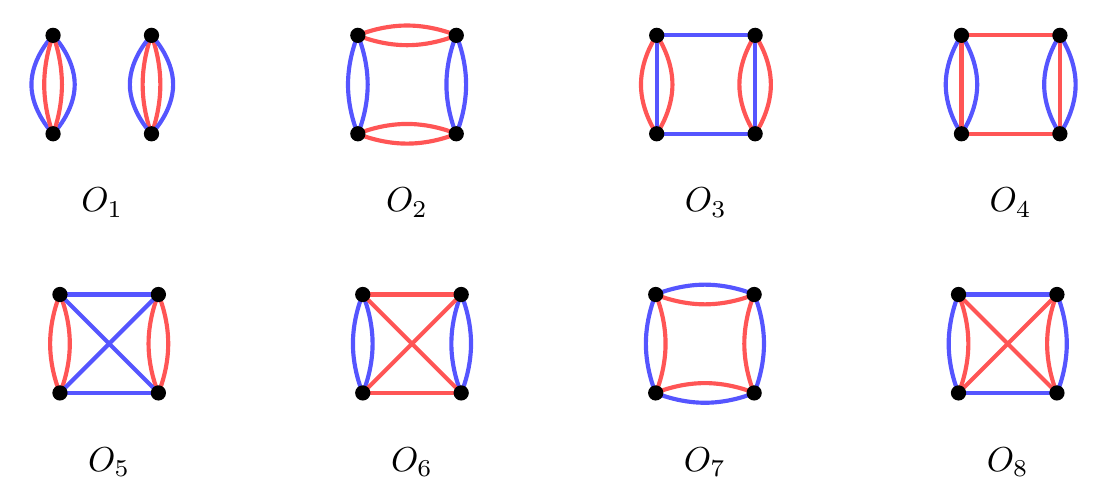}
\caption{Graphical representation of the operators \eqref{eq:oper}}
\label{fig:oper}
\end{figure}
Diagrammatically, these operators can be represented as in Figure \ref{fig:oper}. The beta functions for the coupling constants associated to these operators admit of a perturbative expansion:
\begin{align}
\beta_{g_i}= \mu\frac{dg_i}{d\mu}=-\epsilon g_i+\beta_{g_i}^{(2)}+\mathcal{O}(\| \vec g \|^3)\,,
\end{align}
where $\mu$ is the renormalization group scale. The first term on the RHS represents the naive scalings of the interactions and the terms $\beta_{g_i}^{(2)}$ indicate the lowest-order loop corrections, which are quadratic in the coupling constants. Owing to the large number of operators that mix under the RG flow, analyzing the dynamics of this family of tensor models, even numerically, poses a challenge. Future exploration of exotic QFTs may uncover chaos in smaller systems. But since perturbative beta functions are polynomials with a finite number of roots, by the Poincar\'e-Bendixson theorem chaos cannot occur in continuous theories unless there are at least three coupling constants. We will adopt the perspective of thinking of the beta functions abstractly as an 8-dimensional 2-parameter dynamical system:
\begin{align}
\label{eq:fullsystem}
\frac{dg_i}{dt}=\beta_{g_i}(g_j,N_1,N_2)\,.
\end{align}
Here $t=\log\mu$. Using general formulas for perturbative beta-functions in $4-\epsilon$ dimensions, a straightforward computation yields $\beta_{g_i}^{(2)}$ as functions of the ranks $N_1$ and $N_2$ of the symmetry groups. We list the results in Appendix \ref{appendix:beta}, where explicit formulas for $\beta_{g_i}^{(2)}$ are given. Since the upper and lower indices of the fundamental fields $\phi^{ab}_{\alpha\beta}$ transform in the same kind of representation, in the family of QFTs we are considering, the theory with $O(N_1)\times O(N_2)$ symmetry is the same as the theory with $O(N_2)\times O(N_1)$. This fact is reflected in the beta functions, which are invariant under the simultaneous interchange of $N_1\leftrightarrow N_2$, $g_3\leftrightarrow g_4$, and $g_5 \leftrightarrow g_6$.

When we take $N_1$ and $N_2$ to be integers greater than three, the system of beta-functions describes the behavior of a real unitary quantum field theory, and therefore we expect only regular solutions to the system. The solutions curves are all heteroclinic orbits --- trajectories that connect separate fixed points --- and one can find a Zamolodchikov metric for the system and check that indeed the $a-$theorem is satisfied. In Appendix \ref{appendix:metric} we show the precise form of the metric to leading order in perturbation theory, where the metric is independent of the coupling constants but depends only on $N_1$ and $N_2$. Meanwhile, if $N_1$ or $N_2$ is equal to two or three, the eight operators \eqref{eq:operators} will no longer be independent as there will exist vanishing linear combinations. And if $N_1$ or $N_2$ is equal to one, all eight operators vanish.

In the following, we continue the system to fractional values of $N_1$ and $N_2$. The regime where  either $N_1$ or $N_2$ lies between $-2$ and $3$ is particularly interesting for our purposes, for here the Zamolodchikov metric becomes sign-indefinite so that the RG flow is no longer constrained to be monotonic. It therefore becomes possible for the RG flow to exhibit limit cycles, homoclinic orbits, and chaos. Our focus will be on investigating the presence of chaos. To this end, we look for ZH bifurcations \cite{takens1974singularities,guckenheimer2013nonlinear,kuznetsov2013elements} in the space of $N_1$ and $N_2$. In other words, we are interested in the existence of special fixed points such that
\begin{gather}
\begin{cases}
    \displaystyle    \beta_i\left(g_j,N_1,N_2\right) = 0\,,
    \vspace{2mm}
    \\
    \displaystyle
    \det\left(\frac{\partial \beta_i}{\partial g_j}\right) = 0\,, 
    \vspace{2mm}
    \\
    \displaystyle
    \det\left(\frac{\partial \beta_i}{\partial g_j}-\lambda\right) - \text{has a pair of imaginary roots in $\lambda$}. 
    \end{cases}
    \label{eq:ZHcond}
\end{gather}
The last requirement can be formulated in terms of a closed algebraic equation, but for brevity we will skip it and refer the reader to \cite{Jepsen:2021rhs}. In total, the conditions \eqref{eq:ZHcond} consist of $8+2$ equations, so that generically one expects a discrete set of solutions in the space of eight couplings $g_i$ and two parameters $N_1$ and $N_2$. 
In the vicinity of a ZH point it is possible to make a coupling constant redefinition $\left\{g_i\right\} \to \left\{x,y,z,\mathfrak{g}_i\right\}$, a parameter redefinition $(N_1,N_2) \to (a,b,\eta)$, and a reparametrization $t\rightarrow \tau(t)$ in such a way that the system of differential equations furnished by the beta functions can be brought into the so-called truncated normal form 
parameterized by coordinates $x,y,z$:
\begin{gather}
\begin{cases}
\displaystyle    \frac{dx}{d\tau} = y +\eta x- axz\,, 
\vspace{2mm}
\\
\displaystyle       \frac{dy}{d\tau} = -x+\eta y-ayz\,,
     \label{eq:ZHnormform}
     \vspace{2mm}
    \\
\displaystyle   
    \frac{dz}{d\tau} = -\beta+z^2+b(x^2+y^2)\,.
\end{cases}
\end{gather}
These equations are approximate in that we are omitting terms of cubic and higher order in the coordinates. The parameters $a,b,\eta,\beta$ depend on $N_1$ and $N_2$, with $a$, $\mu$, and $\eta$ being real-valued, while $b=\pm 1$. Right at the ZH point $\beta$ and $\eta$ equal zero. The other coupling constants $\mathfrak{g}_i$ decouple from the rest at this order of expansion. The two-parameter invariant manifold given by $\mathfrak{g}_i=0$ is known as the center manifold. There is a definite procedure for computing the higher order terms on the center manifold, order by order, although the higher order terms never terminate. In Appendix \ref{appendix:normalForm} we present the explicit transformation that puts the beta functions into the truncated normal form.

The system of equations \eqref{eq:ZHnormform} has 6 topologically distinct types of behaviors, depending on whether $b$ equals plus or minus one and on whether $a>-1$, $a\in (-1,0)$, or $a>0$. We are interested in the type with $a,b>0$. The recent result of \cite{baldoma2020hopf} is that in two-parameter dynamical systems with ZH points of this type, there will generically exist nearby parameter values for which the systems exhibit Shilnikov homoclinic orbits and hence chaos. For the system of beta-functions governing the RG flows of the tensor models \eqref{eq:action}, as given in equations (\ref{eq:beta1}-\ref{eq:beta8}) in Appendix \ref{appendix:metric}, it can be verified that there is a ZH point of the particular type in question situated at \sisetup{round-mode=figures,round-precision=4}
\begin{align}\label{eq:ZH}\tag{ZH}
&        \begin{pmatrix*}[r]
        N_1^* \\
        N_2^*
    \end{pmatrix*}= \begin{pmatrix*}[r]
        \num{2.520427996662252} \\
        \num{1.972260741688210} 
    \end{pmatrix*},
    \vspace{3mm}
    \\
    g^\ast \equiv
\big(
      \num{31.01364336772574},\,
      \num{14.90146693485237},\,&
     \num{8.879972225165773},\,
     \num{136.1544856591584},\,
    \num{3.810504206934431},\,
    \num{-143.4849700979769},\,
    \num{-18.64425623143404},\,
    \num{18.15961700372116}
    \big) \eps\,.
    \nonumber
\end{align}
For this ZH point the quadratic normal form coefficients $a,b,\eta,\beta$ in \eqref{eq:ZHnormform} are given to leading order in $\delta N_1 \equiv N_1-N_1^\ast$ and $\delta N_2 \equiv N_2-N_2^\ast$ by 
\begin{align}
 a = 0.8826\,, & \hspace{10mm} b = 1\,,
 \\
 \eta =
 -40.59\,\delta N_1
 -208.3\, \delta N_2\,, &
 \hspace{10mm} \beta = -89.05\,\delta N_1-456.9\, \delta N_2\,.
 \nonumber
\end{align}
If we restrict ourselves to the second order normal form in \eqref{eq:ZHnormform} and disregard the higher-order corrections for a moment, we do not find any homoclinic solutions. Depending on the values of $\delta N_1$ and $\delta N_2$, the value of $\beta$ will be either negative or positive. In the latter case, there is a pair of fixed point located at $x=y = 0$ and $z=\pm \sqrt{\beta}$. This pair of fixed point is connected by heteroclinic solutions. And therefore, at quadratic order in the dynamical variables, we do not get any new phenomena --- the system simply flows from one fixed point to another in a regular way. One of these heteroclinic solutions runs vertically between the two fixed points, the $z$ axis being an invariant manifold of the flow \eqref{eq:ZHnormform}. In the special case when we tune $\delta N_1$ and $\delta N_2$ so as to set $\eta=0$, the remaining heteroclinic solutions admit of a simple description, being orbits that wind around the invariant ellipsoid given by
\begin{align}
z^2+\frac{b}{1+a}(x^2+y^2)=\beta\,.
\end{align}
Once we cease to neglect higher order terms in the normal form, these terms mix the {\it heteroclinic} solutions, and {\it homoclinic} solutions emerge. Essentially, once we perturb the system, an initially vertical orbit flowing from one fixed point will now stray slightly from the $z$-axis and miss the other fixed point and can instead merge onto the erstwhile invariant ellipsoid and flow back to the fixed point whence it originated. Whenever these homoclinic solutions satisfy the Shilnikov condition \eqref{eq:ShilnikovCondition}, and chaos ensues.

\begin{figure}
    \centering
    \ifcompileimages%
  \tikzsetnextfilename{homoclinic_orbit_reduced_system}%
  \input{tikz/homoclinic_orbit_reduced_system}%

    \else
        \includegraphics{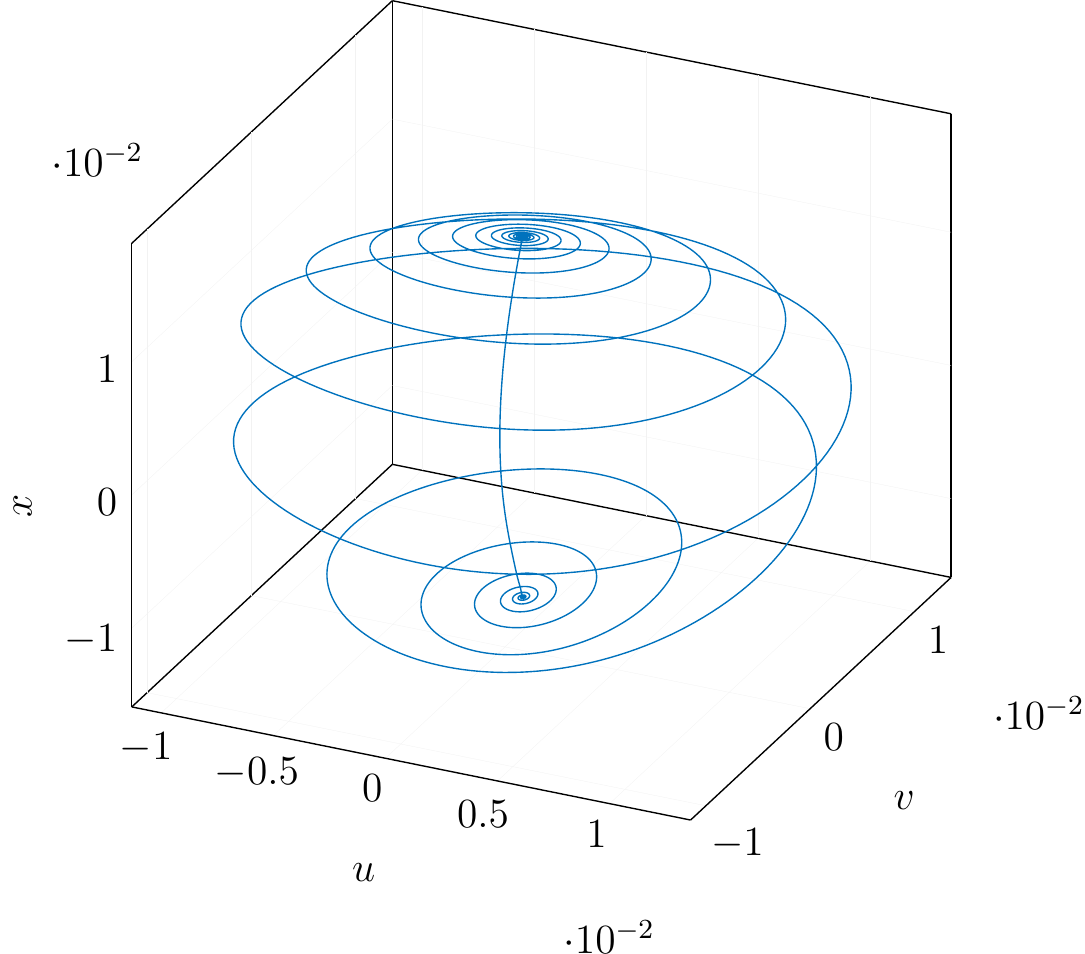}
    \fi
    \caption{Homoclinic orbit to the saddle-focus at $(u,v,x) \approx (0.0008, -0.0001, 0.0152)$ in the RG flow of coupling constants at parameter values $(N_1-N_1^*,N_2-N_2^*)=( \num{7.41174e-8},\num{-5.05368e-7})$. Here $u$, $v$, and $x$ are three independent functions of the couplings $g_i$, with their precise forms given in Appendix \ref{appendix:normalForm}. For computationally reasons, we additionally scaled the variables $u,v$ and $x$ by a factor $10000/(6(32\pi)^2\epsilon)$. In the language of section \ref{sec:Shilnikov}, the fixed point of the homoclinic orbit has eigenvalues $\lambda_{\pm}\approx -0.0020 \pm 0.0307i$  and $\lambda_3 \approx 0.0047$. Thus, the Shilnikov condition is satisfied, and the dynamics near the saddle-focus are complex.
    \label{fig:homoclinic1}}
\end{figure}

\begin{figure}
    \centering
    \ifcompileimages%
  \tikzsetnextfilename{homoclinic_profiles_reduced_system}%
  \input{tikz/homoclinic_profiles_reduced_system}%
 
        \hfill
  \tikzsetnextfilename{homoclinic_profiles_full_system}%
  \input{tikz/homoclinic_profiles_full_system}%

    \else
        \includegraphics{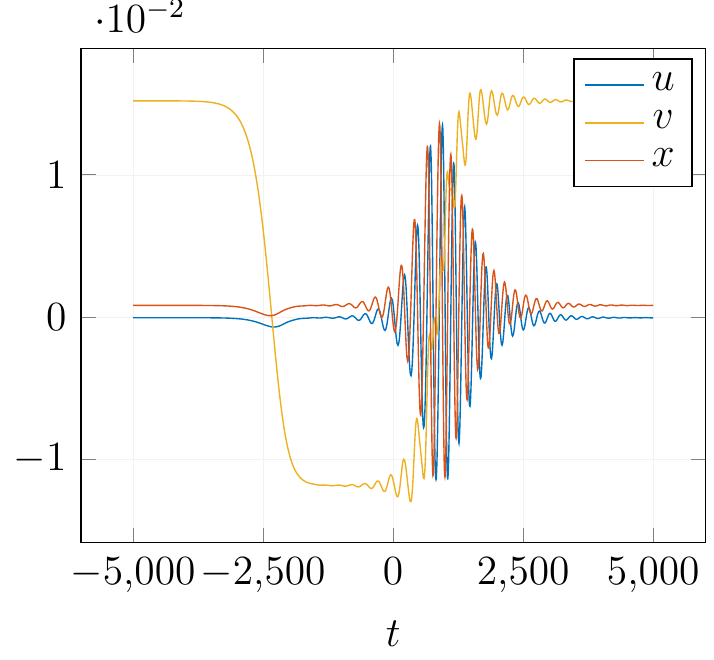}
        \hfill
        \includegraphics{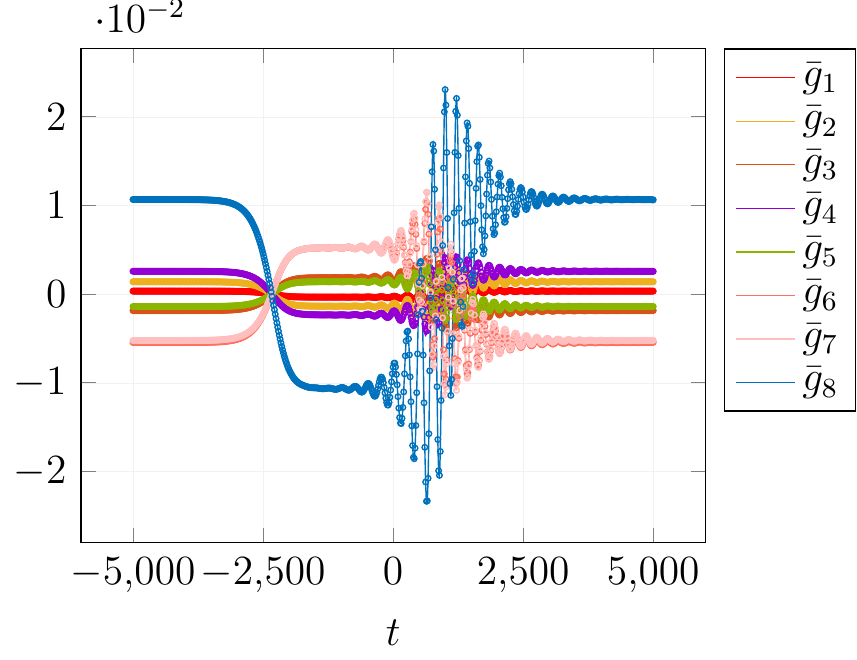}
    \fi
    \caption{In the left panel the profiles of the homoclinic orbit in the reduced system at parameter values $(N_1-N_1^*,N_2-N_2^*)=( \num{7.41174e-8},\num{-5.05368e-7})$, see Figure~\ref{fig:homoclinic1} are shown. In the right panel, a homoclinic solution in the full 8-dimensional space of coupling space at the parameter values $(N_1-N_1^*, N_2-N_2^*)\approx(\num{1.053235334346889e-7}, \num{-6.655194199241109e-7})$. The abscissa represents the span of $t=\log \mu$ truncated to the interval $[-5000, 5000]$, the ordinate indicates values of the eight-coordinates $\bar g$ of the system obtained translating the ZH point to the origin in \eqref{eq:fullsystem} and scaling the resulting coordinates by a factor $10000/(6(32\pi)^2\epsilon)$.
    The solid lines represent the approximated solution, while the circles are the Newton corrected solution. We see that at this scale they are indistinguishable.
     }
    \label{fig:homoclinic1_profiles}
\end{figure}

\begin{figure}
    \centering
    \ifcompileimages%
  \tikzsetnextfilename{wiggling}%
  \input{tikz/wiggling}%

        \hfill
  \tikzsetnextfilename{wigglingRotated}%
  \input{tikz/wigglingRotated}%

    \else
        \includegraphics{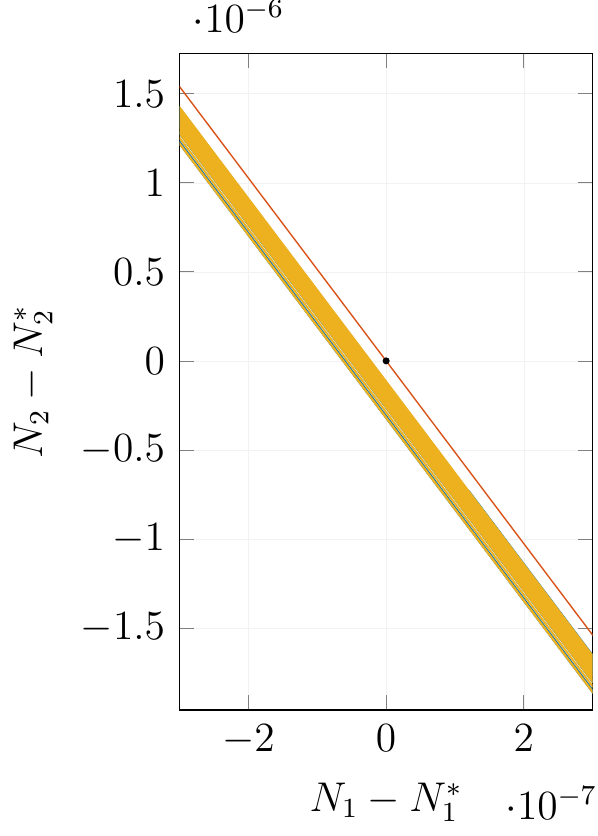}
        \hfill
        \includegraphics{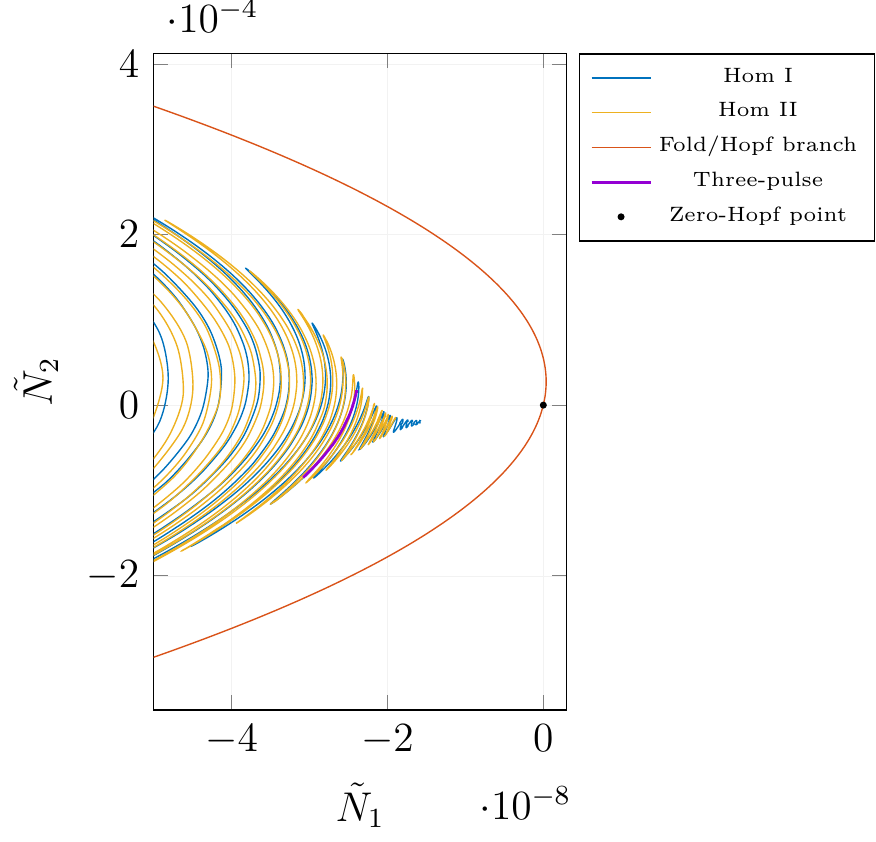}
    \fi
    \caption{The left panel: Entwined wiggling of loci of homoclinic orbits emanating from the zero-Hopf bifurcation point in the shifted parameter space $(N_1-N_1^*,N_2-N_2^*)$. The right panel: Same curves as in the left panel, but rotated clockwise through the angle $0.1925$. Also, a bifurcation curve (purple) of three-pulse Shilnikov homoclinic orbits is shown in the right panel.
    }
    \label{fig:wigcurve}
\end{figure}

For a given ZH point in parameter space, it is a non-trivial task to determine the nearby parameter values for which the dynamical systems exhibit Shilnikov homoclinic orbits. There are no homoclinic asymptotics available as is the case for example in the generic Bogdanov-Takens bifurcation, see \cite{bosschaert2021bt}. Our approach has been to first search for homoclinic solutions in the parameter space of $N_1$ and $N_2$ in the 3-dimensional reduced and truncated system \eqref{eq:reducedsystem} on the center manifold  with cubic terms added in. Subsequently, we were able to uplift these approximate solutions to convergent homoclinic solutions in the full 8-dimensional system.  This was achieved as follows: a homoclinic orbit starts and ends at a fixed point.  The fixed points are not hard to find numerically, since this amounts to setting the beta functions equal to zero and solving the resulting algebraic equations. Once we have a fixed point, we numerically integrate the system, searching for solutions where the one-dimensional unstable manifold of the fixed point is reinjected into its two-dimensional stable manifold. In Figure \ref{fig:homoclinic1} such a solution in the reduced system is shown, while in the left panel of Figure \ref{fig:homoclinic1_profiles} the profiles are given. This solution is then translated to the full system and subsequently, corrected by applying Newton's method to a special boundary value problem, see \cite{DeWitte2012}. The right panel of Figure \ref{fig:homoclinic1_profiles} displays an example of a Shilnikov homoclinic orbit that we located in the vicinity of the ZH bifurcation point \eqref{eq:ZH} in the full system.  This solution in then continued using standard pseudo-arclength continuation, see for example \cite{Beyn2002continuation}.  In the 2-dimensional parameter space spanned by $N_1$ and $N_2$, we observe Shilnikov homoclinic orbits to occur along a one-dimensional subspace that constitutes a wiggling curve emanating from the ZH point, as shown in Figure \ref{fig:wigcurve}. Note that there are two wiggling curves. Indeed, by reversing time, we can use the same method as outlined above to find the second curve of homoclinic orbits. This behavior in parameter space conforms to the general pattern for ZH points of this subtype, as explained in Ref. \cite{champneys2004entwined}, of homoclinic orbits occurring along an oscillating curve in a wedge-shaped region whose thickness decays exponentially as you approach the ZH point. 

We have now ascertained that for a special curve of values for $N_1$ and $N_2$, Shilnikov homoclinic orbits are present in the two-parameter class of dynamical systems \eqref{eq:fullsystem}. This means that the family \eqref{eq:action} of tensor models, there exists a special codimension-one subset of theories whose RG flows contain Shilnikov homoclinic orbits. By the argument reviewed in section \ref{sec:Shilnikov} it follows that for each of these theories, with fixed $N_1$ and $N_2$, the space of coupling constants contains, in addition to a homoclinic orbit, a fractal set of periodic orbits with an infinite range of periodicities. Hence, we conclude that each of these theories exhibits a chaotic RG flow: an arbitrarily tiny change of initial values for the coupling constants can spell the difference between the couplings flowing to a fixed point or them winding endlessly around a loop.

One can object that we are merely working at leading order in perturbation theory in $\epsilon = 4-d$, and that higher loop corrections will modify the RG flow. But the presence of a ZH bifurcation is stable under small deformations, and \cite{baldoma2020hopf} proved rigorously that the presence of Shilnikov homoclinics in a subspace of systems near a ZH point is a generic phenomenon. Since higher-order corrections are suppressed in $\eps$, as long as we take $\eps$ to assume a tiny value we can reliably state that the RG flow is in fact chaotic.

\begin{figure}
    \centering
    \ifcompileimages%
  \tikzsetnextfilename{3pulse}%
  \input{tikz/3pulse}%

        \hfill
  \tikzsetnextfilename{3pulse_full_system}%
  \input{tikz/3pulse_full_system}%

    \else
        \includegraphics{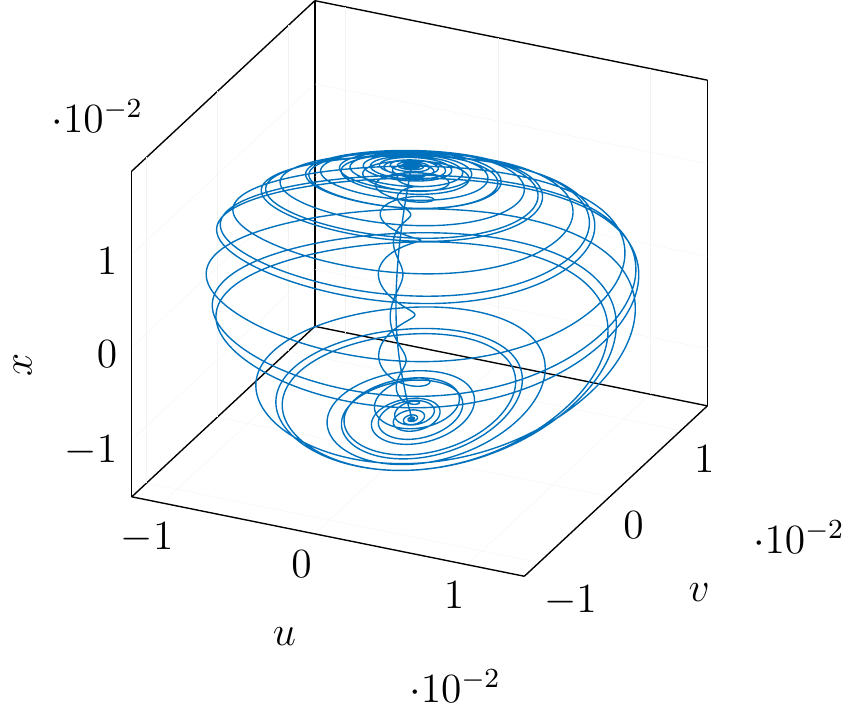}
        \hfill
        \includegraphics{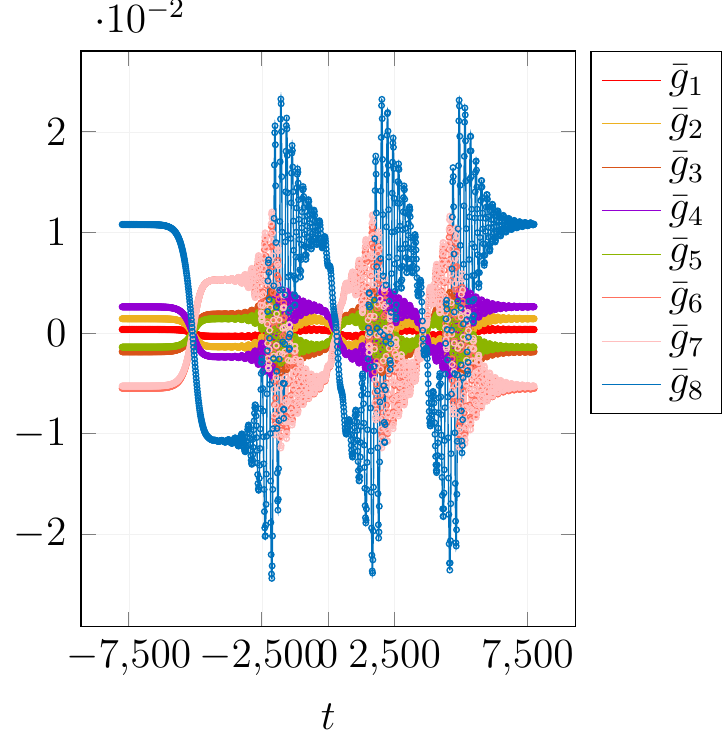}
    \fi
    \caption{The left panel: Three-pulse homoclinic orbit to the saddle-focus at $(u,v,x) \approx (0.0153, 0.0008, -5.0736 \times 10^{-5})$ in the RG flow of coupling constants at parameter values $(N_1-N_1^*,N_2-N_2^*)=( \num{5.144712958706644e-8},\num{-3.915956879261614e-7})$. Here $u$, $v$, and $x$ are three independent functions of the couplings $g_i$, with their precise forms given in Appendix \ref{appendix:normalForm}.
    The right panel: the profile of the three-pulse homoclinic solution in the shifted and scaled 8-dimensional phase-space.
    \label{fig:3pulse}}
\end{figure}

As a symptom of the chaos, it is possible to find a number of interesting or bizarre RG trajectories that arise as we slightly change the initial conditions. We saw above that a homoclinic orbit arises as a heteroclinic orbit is disturbed, such that an orbit originating from one fixed point misses the other fixed point, but winds back to its starting point. By further perturbing this orbit, it is possible to make it miss the starting point and instead flow once more towards the other fixed point, miss it again and wind back whence it originated. Such an orbit is called a 2-pulse homoclinic orbit. Through additional careful deformations, one can find 3-, 4-, 5-, ... pulse homoclinic obits, which densely occupy the neighborhood of the Shilnikov homoclinic orbit. As an example, we display a 3-pulse homoclinic orbit in Figure \ref{fig:3pulse}.

\section{Discussion and Outlook}

In this paper, we have presented examples of QFTs with chaotic RG flows. Usually, in the study of the renormalization group, one considers flows that are non-chaotic and thereby give rise to universality --- the behavior of a system at the macroscopic level is insensitive to the precise microscopic configuration of the system. Imagine that we have two adjacent balls made of the same metal. If we measure their macroscopic properties, such as the specific heat, speed of sound, and density, we would find them to be identical. But if we use an electronic microscope and discern the internal structure of these objects we would find them to be completely different: the distance between the atoms would vary, the number of atoms would be different, the exact distribution of isotopes and defects would differ, and so on. Generally, when an RG flow is non-chaotic, some parameters are irrelevant at large distances so that small alterations at short length scales do not affect the large scale description of the system. But one of the key features of any chaotic mapping is an extreme sensitivity to initial conditions. If we imagine a world with a chaotic renormalization group, macroscopic systems would be sensitive to the properties of each individual atom. In this sense, chaotic RG flows generally, and our example specifically, present a new and peculiar instance of a UV/IR mixing, where the IR behavior is extremely sensitive to the UV properties. For this reason, the study of chaotic RG flows and the demarcation of the instances where they occur pose an interesting problem. In the examples we considered, it was necessary to relax the condition of unitary, but such theories can nonetheless be studied on a lattice or find realization in cold atom experiments. And so to the question raised in the title of the paper \cite{morozov2003can}, we respond that {\it yes, RG flows can indeed end up in a total mess.} 

\begin{figure}
\ifcompileimages
  \tikzsetnextfilename{simulate_2_torus}%
  \input{tikz/simulate_2_torus}%

    \hfill
  \tikzsetnextfilename{torus}%
  \input{tikz/torus}%

\else
    \includegraphics[width=0.45\textwidth]{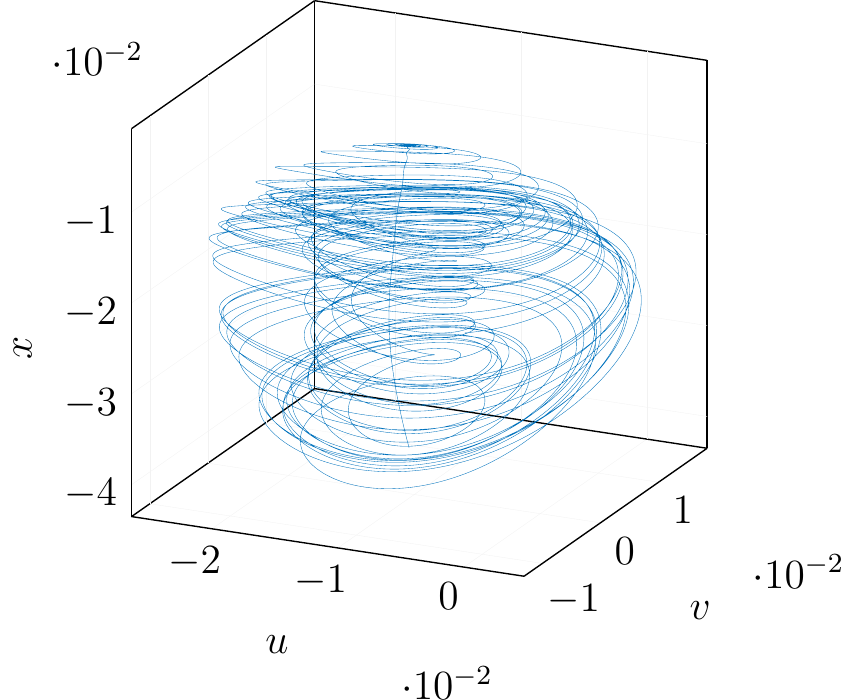}%
    \hfill
    \includegraphics[width=0.45\textwidth]{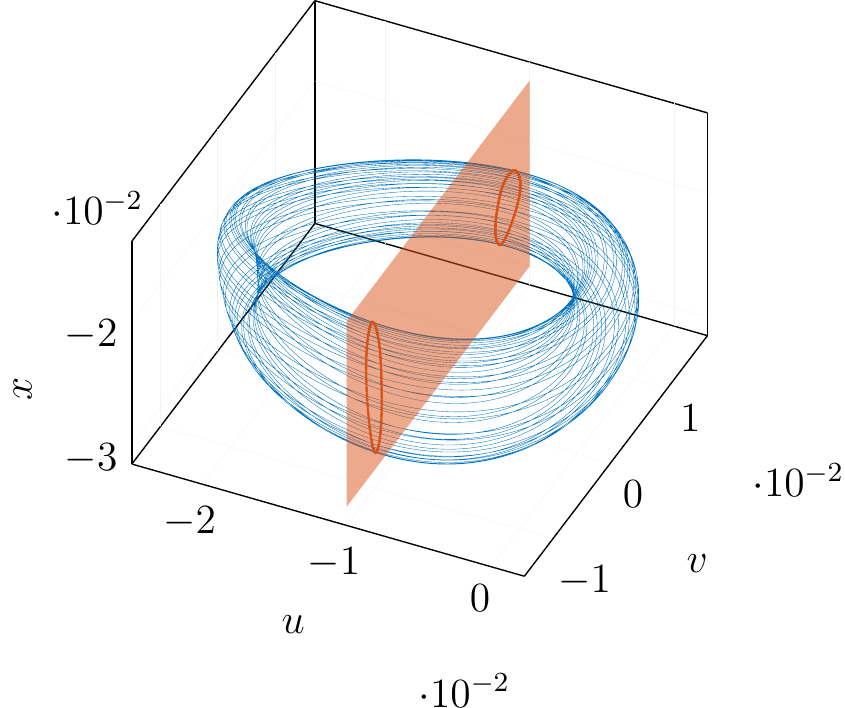}
\fi
\caption{In the left panel: A spiral attractor-like solution of the reduced system \eqref{eq:reducedsystem} obtained by integrating backwards in time starting in the one-dimensional unstable manifold of the equilibrium located at $(u,v,x)\approx (\num{-0.02473081349681433},\num{-0.02190477549603949},\num{0.0014168404408595782})$ with parameter values $(N_1-N_1^*, N_2-N_2^*) = (\num{-3.4894952252725343e-6}, \num{1.7741178541515484e-5})$.
The solution convergence to a two-dimensional torus, shown in the right panel.
\label{fig:2intersol}}
\end{figure}

Having established the existence of QFTs with chaotic RG flows, it is reasonable to expect these theories to manifest period-doubling cascades, Feigenbaum scaling, strange attractors and other interesting phenomena. For the tensor model \eqref{eq:action} studied in this paper, it is a formidable numerical challenge to reliably investigate exotic RG trajectories because the flow equations are 8-dimensional. But as an approximate probe of the behavior of the full system, we can study the truncated flow equations describing the reduced system on the center manifold in the hope that these solution curves can be uplifted to the full system, like we were able to do for the Shilnikov homoclinic orbits. Generally, when searching for exotic trajectories in the reduced system, we observe three kinds of solution curves:
\begin{enumerate}
\item The solution grows without bound.
\item The solution converges to a periodic cycle.
\item The solution converges to a torus.
\end{enumerate}
An example of the third scenario is depicted in Figure \ref{fig:2intersol}.

It is a well-known and much appreciated fact that Lagrangians, in furnishing virtually all the essential information of the QFTs they represent, provide a beautifully compact way of encoding a tremendous amount of information. We leave it as a question for future work whether this information includes the geometry of fractal patterns.

\section{Acknowledgment}

We are grateful to Igor R. Klebanov for insightful discussions and suggestions throughout the project. We are also grateful to A. Gorsky, A.Morozov, Yu.\ A. Kuznetsov, A. Milekhin, Y. Oz,  A. Polyakov, Y. Wang, S. Dubovsky and V. Rosenhaus for valuable discussions and comments. 
FKP was supported by the Russian Science Foundation (Grant No. 20-71-10073).

\appendix

\section{General Discussion of Chaos}

\label{appendix:chaos}

 \epigraph{ “The least initial deviation from the truth is multiplied later a thousandfold”}{{\it Aristotle} \cite{Aristotle}}

It is rather hard to formulate a precise mathematical definition of chaos, and for this reason one sometimes encounters definitions of chaos of a speculative or even philosophical nature. From a pedestrian point of view, if by "chaos" we have in mind something unpredictable, the equations \eqref{eq:dynsys} should not contain any chaos: if we know the initial state of a system, we can evolve it forward in time and predict all future states exactly. Due to the uniqueness and existence theorems for systems of differential equations, we know that systems subject to the same initial conditions will follow the same evolution. Therefore, we would not expect randomness or "chaos" in our dynamical system \eqref{eq:dynsys}. The imaginary intellect whose full knowledge of the present universe allows it to know with certitude all future course of history is sometimes referred to as Laplace's demon.  But of course in actuality, we never know exactly the initial state of a system, and one of the most intuitive features of chaos is the dependence on initial conditions.  Due to finite experimental precision and unknown aspects of the system, we can at best know the initial state to within a finite degree of precision. Therefore, one could reasonably say that we should study the dynamics of an $\eps$-domain of the initial state of the system $U_\eps(g^0) = \left\{g_j: \left|\left| g_i-g_i^0\right| \right| < \eps \right\}$. And if the evolution of this $\eps$-domain leaves the domain still of finite size or even shrinking in time, we would expect the system in question not to be chaotic. For any initial point $g_* \in U_\eps(g^0)$, we have some information about the future of its trajectory, namely, $g_*(t) \in U_\eps(g^0(t))$. If this region $U_\eps(g^0(t))$ grows larger and larger in time and even covers the whole phase space, then we would not be able to make any predictions as to the state of the dynamical system far into the future. Such a system we could say is chaotic for which a small perturbation of the initial state drastically changes the behavior of the system, as in the famous butterfly effect. In classical and quantum mechanics, one usually associates such a behavior where the initial $\eps$-domain $U_\eps$ grows uncontrollably large by the appearance of a non-zero Lyapunov exponent $\lambda_L$:
\begin{gather}
\left|\left|\delta q(t)\right|\right| \sim \left|\left|\delta q(t=0)\right|\right| \exp\left[\lambda_L t \right]\,. \label{eq:LyapExp}
\end{gather}
A small perturbation of the initial state leads to an exponential error in the later observables. This property of dynamical systems is sometimes referred to as strong sensitive dependence.
Lyapunov exponents have been computed for some famous models like SYK \cite{maldacena2016remarks}. The existence of a positive Lyapunov exponent is suggestive of chaos, but it is not a sufficient condition. Indeed, consider the following dynamical system
\begin{gather}
    g(x) = 2 x\,, \quad g^n(x) = 2^n x\,. \label{eq:Lyapcounterex}
\end{gather}
Obviously this system has a positive Lyapunov exponent, but it is nevertheless not considered chaotic. In the setting of RG flows, local exponential behavior as in \eqref{eq:Lyapcounterex} is commonplace for systems that exhibit universality and eventually flow to gapped theories or CFTs. Another issue with the condition \eqref{eq:LyapExp} as a criterion for chaos is that for systems with compact phase spaces, it cannot be satisfied for all times. Moreover, the existence of a global Lyapunov exponent can in some cases be an observer-dependent phenomenon, for example a system with a non-zero Lyapunov exponent in Minkowski space can have a vanishing Lyapunov exponent in Rindler space \cite{zheng2003observer}. So we see that the condition \eqref{eq:LyapExp} is neither a necessary nor a sufficient diagnostic for chaos. 

To resolve the above issues, one of the most dominant ideas is to replace the condition of strong sensitivity dependence with a weak sensitivity condition:
\begin{enumerate}
    \item[1.] The distance between the initial conditions grows with time: $\left|\left|\delta q(t_2)\right|\right| > \left|\left|\delta q(t_1)\right|\right|$ if $t_2>t_1$.
\end{enumerate}
and to add the following two additional conditions \cite{devaney1989introduction}:
 \begin{enumerate}
     \item[2.] There is a dense set of periodic orbits in the phase space.
     \item[3.] The dynamical system is topologically transitive on the phase space, meaning that if we take any two open regions $U$ and $W$ of phase space, then the evolution of $U$ along the dynamical system would bring it to intersect with $W$ : $g^t(U) \cap W \neq \emptyset$.
 \end{enumerate}
 In the examples of chaotic systems provided in the main text, one can check that these three conditions are indeed satisfied. 
To systems with RG flows that are chaotic according to these conditions, the concept of universality does not apply. Arbitrarily small changes in the initial conditions, say any slight tweak of the coupling constants in the UV theory or the replacement of a square lattice with a triangular lattice, drastically alter the behavior of the system and can spell the difference between periodic orbits and ergodic motion in phase space.

\section{The Two-Loop Beta Functions and the Zamolodchikov Metric}
By the use of the general formulas for any marginal scalar field theory in four dimensions \cite{Jack:1990eb}, one finds that the beta functions for the eight couplings of the family of the models \eqref{eq:action} are given by
\label{appendix:beta}
\label{appendix:metric}

\begin{align}
\beta_{g_1}^{(2)}
=&\,
\frac{1}{16\pi^2}
\bigg(
 \frac{N_1^2 N_2^2 - N_1^2 N_2 - N_1 N_2^2+ N_1 N_2 +32}{12} g_1^2
  + \frac{ N_1^2+ N_2^2- N_1 - N_2 +2}{3} g_1 g_2 
   \nonumber\\ &
  + \frac{N_1^2 N_2- N_1^2 - N_1 N_2+ N_1 + 2 N_2 }{6} g_1 g_3
  +  \frac{N_1 N_2^2- N_1 N_2 - N_2^2+2 N_1 + N_2  }{6} g_1 g_4 
   \nonumber\\ &
  + \frac{N_1^2-N_1 + 4 N_2-4}{6}g_1 g_5
  +  \frac{N_2^2 + 4 N_1 - N_2 -4}{6} g_1 g_6 
  +  \frac{2 N_1 N_2-2N_1-2N_2+3}{6}g_1 g_7
     \nonumber\\ &
  + \frac{N_1 N_2-1}{6} g_1 g_8 
  + g_2^2 
  + \frac{N_2-1}{3} g_2 g_3
  + \frac{N_1-1}{3} g_2 g_4 
  + \frac{g_2 g_5}{3} 
  + \frac{g_2 g_6}{3}  
  + \frac{N_1^2-N_1+6}{24} g_3^2
       \nonumber\\ &
  + \frac{(N_1-1)(N_2-1)}{6} g_3 g_4 
  + \frac{g_3 g_5}{3}
  + \frac{N_2-1}{6} g_3 g_6  
  + \frac{N_1-1}{6} g_3 g_7
  + \frac{N_1}{12}g_3 g_8
       \nonumber\\ &
 +  \frac{N_2^2- N_2 +6}{24} g_4^2 
  + \frac{N_1-1}{6} g_4 g_5 
  + \frac{g_4 g_6}{3} 
  + \frac{N_2-1}{6} g_4 g_7
  + \frac{N_2}{12}g_4 g_8
    + \frac{g_5 g_6}{6} 
           \nonumber\\ &
  + \frac{g_7^2}{8} 
  + \frac{g_7 g_8}{12} 
  + \frac{g_8^2}{48}
\bigg)\label{eq:beta1}
\\ \nonumber
\\
\beta_{g_2}^{(2)}
=&\,
\frac{1}{16\pi^2}
\bigg(
4 g_1 g_2 
+ \frac{N_1^2+N_2^2-N_1-N_2+8}{6} g_2^2
+ \frac{N_2}{3}g_2 g_3  
+ \frac{N_1}{3}g_2 g_4  
+ \frac{2(N_2-1)}{3} g_2 g_5  
  \nonumber\\ &
+ \frac{2(N_1-1)}{3} g_2 g_6  
+ \frac{1}{6}g_3^2 
+ \frac{2}{3} g_3 g_5 
+ \frac{1}{6}g_4^2 
+ \frac{2}{3} g_4 g_6
+ \frac{N_1^2-N_1+4}{12} g_5^2 
+ \frac{1}{6}g_5 g_7 
  \nonumber\\ &
+ \frac{N_1-1}{6} g_5 g_8
+ \frac{N_2^2-M+4}{12} g_6^2 
+ \frac{1}{6}g_6 g_7 
+ \frac{N_2-1}{6} g_6 g_8 
+ \frac{1}{24}g_7^2 
+ \frac{1}{12}g_8^2 
\bigg)
\\ \nonumber
\\
\beta_{g_3}^{(2)}
=&\,
\frac{1}{16\pi^2}
\bigg(
4 g_1 g_3 
+ \frac{N_1^2-N_1+6}{3} g_2 g_3 
+ \frac{8}{3} g_2 g_5 
+ \frac{2(N_1-1)}{3} g_2 g_7 
+ \frac{N_1}{3}g_2 g_8  
  \nonumber\\ &
+ \frac{N_1^2 N_2-2N_1^2-N_1 N_2+2N_1+6 N_2-4}{24} g_3^2 
+ \frac{N_1}{3}g_3 g_4  
+ \frac{N_1^2-N_1+2 N_2}{6} g_3 g_5  
  \nonumber\\ &
+ \frac{2(N_1-1)}{3} g_3 g_6
+\frac{N_1 N_2-2N_1- N_2+3}{6} g_3 g_7
+ \frac{N_1 N_2-2}{12} g_3 g_8
+ \frac{2}{3} g_4 g_7 
+ \frac{1}{3}g_4 g_8 
  \nonumber\\ &
  + \frac{N_2}{3}g_5^2 
  + \frac{N_1-1}{3} g_5 g_7 
+ \frac{N_1}{6}g_5 g_8  
+ \frac{2}{3} g_6 g_7
+ \frac{1}{3}g_6 g_8 
+ \frac{N_2-3}{12} g_7^2 
+ \frac{N_2-1}{12} g_7 g_8 
+ \frac{N_2}{24} g_8^2 
\bigg)
\\ \nonumber
\\
\beta_{g_4}^{(2)}
=&\,
\frac{1}{16\pi^2}
\bigg(
4 g_1 g_4 
+ \frac{N_2^2-N_2+6}{3} g_2 g_4 
+ \frac{8}{3} g_2 g_6 
+ \frac{2(N_2-1)}{3} g_2 g_7 
+ \frac{N_2}{3}g_2 g_8 
+ \frac{N_2}{3}g_3 g_4 
 \nonumber\\ &
+ \frac{2}{3} g_3 g_7 
+ \frac{1}{3}g_3 g_8 
+ \frac{N_1 N_2^2- N_1 N_2-2N_2^2+6 N_1+2N_2-4}{24} g_4^2 
+ \frac{2(N_2-1)}{3} g_4 g_5
 \nonumber\\ &
+ \frac{2N_1-N_2+N_2^2}{6} g_4 g_6 
+ \frac{N_1 N_2-N_1-2N_2+3}{6} g_4 g_7 
+ \frac{N_1 N_2-2}{12} g_4 g_8 
+ \frac{2}{3} g_5 g_7 
 \nonumber\\ &
 + \frac{1}{3}g_5 g_8 
+ \frac{N_1}{3}g_6^2 
+  \frac{(N_2-1)}{3} g_6 g_7 
+ \frac{N_2}{6} g_6 g_8 
+ \frac{N_1-3}{12} g_7^2 
+ \frac{N_1+1}{12} g_7 g_8 
+ \frac{N_1}{24}g_8^2 
 \bigg)
\\ \nonumber
\\
\beta_{g_5}^{(2)}
=&\,
\frac{1}{16\pi^2}
\bigg(
 4 g_1 g_5 
+ \frac{4}{3} g_2 g_3 
+ \frac{N_1^2-N_1+2}{3} g_2 g_5  
+ \frac{1}{3}g_2 g_7 
+ \frac{N_1-1}{3} g_2 g_8 
+ \frac{1}{6}g_3^2 
+ \frac{N_1}{3}g_4 g_5  
 \nonumber\\ &
+ \frac{1}{3}g_4 g_8 
- \frac{N_1(N_1-1)}{12} g_5^2 
+ \frac{2(N_1-1)}{3} g_5 g_6 
- \frac{1}{6}g_5 g_7 
-\frac{N_1-1}{6} g_5 g_8 
+ \frac{1}{3} g_6 g_8 
+ \frac{1}{12}g_7 g_8
\nonumber \\ & 
+ \frac{N_2-4}{48} g_8^2 
+ \frac{N_2-2}{3} g_3 g_5
\bigg)
\\ \nonumber
\\
\beta_{g_6}^{(2)}
=&\,
\frac{1}{16\pi^2}
\bigg(
4 g_1 g_6 
+ \frac{4}{3} g_2 g_4 
+ \frac{N_2^2-N_2+2}{3} g_2 g_6
+ \frac{1}{3}g_2 g_7 
+ \frac{N_2-1}{3} g_2 g_8 
+ \frac{N_2}{3} g_3 g_6 
+ \frac{1}{3}g_3 g_8 
\nonumber \\ & 
+ \frac{1}{6}g_4^2 
+ \frac{N_1-2}{3} g_4 g_6 
+ \frac{2(N_2-1)}{3} g_5 g_6 
- \frac{N_2(N_2-1)}{12} g_6^2
+ \frac{1}{3}g_5 g_8 
- \frac{1}{6}g_6 g_7 
\nonumber \\ & 
- \frac{N_2-1}{6} g_6 g_8 
+ \frac{1}{12}g_7 g_8 
+ \frac{N_1-4}{48} g_8^2 
\bigg)
\\ \nonumber
\\
\beta_{g_7}^{(2)}
=&\,
\frac{1}{16\pi^2}
\bigg(
4 g_1 g_7 
+ \frac{2}{3} g_2 g_7 
+ \frac{2}{3} g_2 g_8 
+ \frac{4}{3} g_3 g_4 
+ \frac{N_2-2}{3} g_3 g_7  
+ \frac{1}{3} g_3 g_8 
+ \frac{N_1-2}{3} g_4 g_7 
+ \frac{1}{3}g_4 g_8 
\nonumber \\ & 
+ \frac{8}{3} g_5 g_6 
+ \frac{N_2-1}{3} g_5 g_8 
+ \frac{N_1-1}{3} g_6 g_8 
+ \frac{N_1 N_2-2N_1-2N_2+7}{12} g_7^2 
\nonumber \\ & 
+ \frac{N_1+N_2-3}{12} g_7 g_8 
+ \frac{N_1 N_2}{48} g_8^2 
\bigg) 
\\ \nonumber
\\ 
\beta_{g_8}^{(2)}
=&\,
\frac{1}{16\pi^2}
\bigg(
4 g_1 g_8 
+ \frac{4}{3} g_2 g_7 
+ \frac{4}{3} g_2 g_8
+ \frac{4}{3} g_3 g_4 
+ \frac{8}{3} g_3 g_6 
+ \frac{2}{3} g_3 g_7 
+ \frac{N_2-1}{3} g_3 g_8
+ \frac{8}{3} g_4 g_5 
\nonumber \\ & 
+ \frac{2}{3} g_4 g_7 
+ \frac{N_1-1}{3} g_4 g_8 
+ \frac{2(N_2-1)}{3} g_5 g_7 
+ \frac{N_2-1}{3} g_5 g_8 
+ \frac{2(N_1-1)}{3} g_6 g_7 
\nonumber \\ & 
+ \frac{N_1-1}{3} g_6 g_8 
+ \frac{1}{6}g_7^2 
+ \frac{N_1 N_2-N_1-N_2-3}{12} g_7 g_8
+ \frac{1}{12}g_8^2 
\bigg) \label{eq:beta8}
\end{align}
These beta-functions are gradient $\beta_i = G_{ij}\partial^j A$. The Zamolodchikov metric $G_{ij}$ that governs the flow has, up to a convention-dependent overall normalization, the following components:
\begin{align}
G_{11}=\,&4 (8 + N_1 N_2 - N_1^2 N_2 - N_1 N_2^2 + N_1^2 N_2^2)
\nonumber \\
G_{12}=\,&8 (2 - N_1 + N_1^2 - N_2 + N_2^2)
\nonumber \\
G_{13}=\,&4 (N_1 - N_1^2 + 2 N_2 - N_1 N_2 + N_1^2 N_2)
\nonumber \\
G_{14}=\,&4 (2 N_1 + N_2 - N_1 N_2 - N_2^2 + N_1 N_2^2)
\nonumber \\
G_{15}=\,&4 (-4 - N_1 + N_1^2 + 4 N_2)
\nonumber \\
G_{16}=\,&4 (-4 + 4 N_1 - N_2 + N_2^2)
\nonumber \\
G_{17}=\,&4 (3 - 2 N_1 - 2 N_2 + 2 N_1 N_2)
\nonumber \\
G_{18}=\,& 4 (-1 + N_1 N_2)
\nonumber \\
G_{22}=\,&2 (12 - 2 N_1 + 2 N_1^2 - 2 N_2 + N_1 N_2 - N_1^2 N_2 + 2 N_2^2 - N_1 N_2^2 + N_1^2 N_2^2)
\nonumber \\
G_{23}=\,&2 (-4 + 6 N_2 - N_1 N_2 + N_1^2 N_2)
\nonumber \\
G_{24}=\,&2 (-4 + 6 N_1 - N_1 N_2 + N_1 N_2^2)
\nonumber \\
G_{25}=\,&4 (N_1 - N_1^2 + 2 N_2 - N_1 N_2 + N_1^2 N_2)
\nonumber \\
G_{26}=\,&4 (2 N_1 + N_2 - N_1 N_2 - N_2^2 + N_1 N_2^2)
\nonumber \\
G_{27}=\,&4 (-1 + N_1 N_2)
\nonumber \\
G_{28}=\,&2 (2 - 2 N_1 - 2 N_2 + 3 N_1 N_2)
\nonumber \\
G_{33}=\,&\frac{1}{2} (8 - 4 N_1 + 4 N_1^2 + 2 N_1 N_2 - 2 N_1^2 N_2 + 2 N_2^2 - N_1 N_2^2 + N_1^2 N_2^2)
\nonumber \\
G_{34}=\,&2 (2 - 2 N_1 - 2 N_2 + 3 N_1 N_2)
\nonumber \\
G_{35}=\,&8 - 4 N_2 - N_1 N_2 + N_1^2 N_2 + 2 N_2^2
\nonumber \\
G_{36}=\,&4 (-1 + N_1 N_2)
\nonumber \\
G_{37}=\,&-4 + 4 N_1 + 3 N_2 - 2 N_1 N_2 - N_2^2 + N_1 N_2^2
\nonumber \\
G_{38}=\,&\frac{1}{2} (4 N_1 - 2 N_2 + N_1 N_2^2)
\nonumber \\
G_{44}=\,&\frac{1}{2} (8 + 2 N_1^2 - 4 N_2 + 2 N_1 N_2 - N_1^2 N_2 + 4 N_2^2 - 2 N_1 N_2^2 + N_1^2 N_2^2)
\nonumber \\
G_{45}=\,&4 (-1 + N_1 N_2)
\nonumber \\
G_{46}=\,&8 - 4 N_1 + 2 N_1^2 - N_1 N_2 + N_1 N_2^2
\nonumber \\
G_{47}=\,&-4 + 3 N_1 - N_1^2 + 4 N_2 - 2 N_1 N_2 + N_1^2 N_2
\nonumber \\
G_{48}=\,&\frac{1}{2} (-2 N_1 + 4 N_2 + N_1^2 N_2),
\nonumber \\
G_{55}=\,&-4 N_1 + 4 N_1^2 + 4 N_2 + 3 N_1 N_2 - 3 N_1^2 N_2 - N_1 N_2^2 + N_1^2 N_2^2
\nonumber \\
G_{56}=\,&4 (3 - 2 N_1 - 2 N_2 + 2 N_1 N_2)
\nonumber \\
G_{57}=\,&4 - 5 N_2 + 2 N_1 N_2 + N_2^2
\nonumber \\
G_{58}=\,&-4 + 4 N_1 + 3 N_2 - 2 N_1 N_2 - N_2^2 + N_1 N_2^2
\nonumber \\
G_{66}=\,&4 N_1 - 4 N_2 + 3 N_1 N_2 - N_1^2 N_2 + 4 N_2^2 - 3 N_1 N_2^2 + N_1^2 N_2^2
\nonumber \\
G_{67}=\,&4 - 5 N_1 + N_1^2 + 2 N_1 N_2
\nonumber \\
G_{68}=\,&-4 + 3 N_1 - N_1^2 + 4 N_2 - 2 N_1 N_2 + N_1^2 N_2
\nonumber \\
G_{77}=\,&\frac{1}{4} (16 - 12 N_1 + 4 N_1^2 - 12 N_2 + 11 N_1 N_2 - 3 N_1^2 N_2 + 4 N_2^2 - 3 N_1 N_2^2 + 
   N_1^2 N_2^2)
\nonumber \\
G_{78}=\,&\frac{1}{4} (4 N_1 + 4 N_2 - 5 N_1 N_2 + N_1^2 N_2 + N_1 N_2^2)
\nonumber \\
G_{88}=\,&\frac{1}{8} (16 - 8 N_1 - 8 N_2 + 4 N_1^2  + 6 N_1 N_2 + 4 N_2^2 - 2 N_1^2 N_2  - 2 N_2 N_2^2 + 
   N_1^2 N_2^2)
\end{align}
The determinant of this metric is given by
\begin{align*}
\det G = \frac{1}{16} (N_1-3)^4 (N_1-2 )^6 (N_1+1)^2 (N_1+2)^4 (N_2-3)^4 (N_2-2)^6 (N_2+1)^2 (N_2+2)^4.
\end{align*}

\noindent When $-2 < N_1 < 3$ or $-2 < N_2 < 3$, the metric has both positive and negative eigenvalues. If $N_1$ and $N_2$ are each greater than 3 or less then $-2$, it is positive-definite. In figure \ref{Eigenvalues} we plot the eight eigenvalues of the Zamolodchikov metric for various values of $N_1$ and $N_2$. In general it is not possible to provide closed-form expressions for the eigenvalues, since they are roots of an eigth-order polynomial. An exception occurs at $N_1=2$, where six eigenvalues equal zero, while the remaining two are given by
\begin{align}
\lambda_\pm = \frac{100-25N_2+25N_2^2}{2}
\pm \frac{5}{2}\sqrt{208-328N_2+337N_2^2-18N_2^3+9N_2^4}\,.
\end{align}

\section{Reduced Equations on the Central Manifold}
\label{appendix:normalForm}
In this appendix we derive the reduced equations on the parameter-dependent center manifold of the system of ordinary differential equations given by
\begin{equation} \label{eq:ODE}
    \dot g_i = \beta_i(\vec{g}, \vec{\alpha}) = -g_i + \beta^{(2)}_i(\vec g, \vec \alpha)\,,
\end{equation}
at the (numerically derived) zero-Hopf point located at $\alpha^* = (N_1^*, N_2^*)$, see \eqref{eq:ZH}. Here the functions $\beta^{(2)}$ are listed in equations (\ref{eq:beta1}-\ref{eq:beta8}) and depend on the vector-valued function $\vec{g} \colon \mathbb R \rightarrow \mathbb R^8$ and the vector $\vec{\alpha}\textbf{} = (N_1,N_2) \in \mathbb R^2$. For brevity of notation, we use units where $\eps=1$. The analysis is general and applies to any 2-parameter dynamical system exhibiting a ZH bifurcation. For readability, all calculations will be rounded up to four figures 
from here on out, although to reproduce the figures in section \ref{sec:QFT} it is necessary to work at significantly higher precision. The derivation of the normal form is also carefully worked out in the textbook \cite{Guao2013BifurcationTheory}, see also \cite{kuznetsov2013elements}.

First, we translate the bifurcation point to the origin by introducing
new variables $\bar g \equiv \vec g - g^\ast$ and parameters $\bar \alpha \equiv \vec \alpha -
\alpha^\ast$. We immediately drop the bars again for readability. Thus, the
ZH point is now located at $(\vec{g}, \vec{\alpha}) = (0,0) \in \mathbb
R^{8 + 2}$. The right-hand side of \eqref{eq:ODE} can be expanded as
\begin{align}
    \beta(\vec{g}, \vec{\alpha}) =& A\vec g + B(\vec{g},\vec{\alpha}) = \mathcal \beta_{0,1} (\vec{\alpha}) + \frac12 \mathcal \beta_{2,0}(\vec g,\vec g) + \mathcal \beta_{1,1}(\vec g,\vec \alpha) + \frac12 \mathcal \beta_{0,2}(\vec \alpha,\vec \alpha)  + \frac12 \mathcal \beta_{2,1}(\vec g,\vec g,\vec \alpha) + \nonumber \\
                   &\frac12 \mathcal \beta_{1,2}(\vec g, \vec \alpha, \vec \alpha) + \frac16 \mathcal \beta_{0,3}(\vec \alpha,\vec \alpha,\vec \alpha) + \cdots \label{eq:Fexpansion}
\end{align}
where $A_{ij} =\partial_{g_i} \beta^j(\vec g,\vec \alpha)$ is the Jacobian of $\beta^j(\vec g,\vec \alpha)$ at the ZH point, 
the notation $\beta_{i,j}$ means that we expand to the $i^{\text{th}}$ order in coupling constant and to the $j^{\text{th}}$ order in parameters
\begin{multline}
   \mathcal \beta_{i,j}(\vec{g}_1, \vec{g}_2, \cdots, \vec{g}_i, \vec{\alpha}_1, \vec{\alpha}_2 \cdots, \vec{\alpha}_j)  \\
    = \frac{\partial^i}{\partial t_1 \partial t_2 \dots \partial t_i}
      \frac{\partial^j}{\partial s_1 \partial s_2 \dots \partial s_j}
        \left. \beta\left(\sum_{r=0}^i t_r \vec{g}_r,  \sum_{r=0}^j s_r \vec{\alpha}_r\right) 
            \right\vert_{t_r=s_r=0}.
\end{multline}
Note that we did not include the multilinear form $\mathcal \beta_{3,0}$ in
\eqref{eq:Fexpansion} since it is identically zero in the perturbative approximation we are working at, where $\beta(g,\alpha)$ is only computed up to second order in the coupling constants. Now we truncate the expansion \eqref{eq:Fexpansion} at the third order in the sum $I=i+j\leq 3$. The reason we perform this double expansion and truncate parameters and couplings together is that we are interested in RG trajectories near the ZH point that do not run away but exhibit fluctuations in coupling space that scale with the magnitude of $\vec{\alpha}$.

To confirm that the point under consideration is indeed a ZH point we
inspect the eigenvalues of the linearization of \eqref{eq:ODE} at
\eqref{eq:ZH}, i.e., the eigenvalues of the matrix $A$. This set can be divided
into three parts
\begin{align*}
    \sigma_s &= 
\left\{
    \num{-0.48779687722209303}, \;
    \num{-0.22070478585058576}, \;
    \num{-0.11374063283553064}
\right\}, \,
    \sigma_c = 
\left\{
   \pm 0.02756 i,\;
    0
\right\}, \,
    \sigma_u = 
\left\{
    \num{0.40013960733401366}, \;
    1
\right\},
\end{align*}
corresponding to the UV-stable, center, and UV-unstable eigenspaces of $A$, respectively. The eigenvalue $1$ is exact at second order in perturbation theory and owes to the homogeneity of $\beta+g$ \cite{Jepsen:2020czw}. 
As a shorthand we introduce $\omega = 0.02756$.
Then, since the eigenvalues in $\sigma_c$ are non-degenerate or simple, there are eigenvectors $p_0\in
\mathbb R^8$ and $p_1 \in \mathbb C^8$, and adjoint eigenvectors $q_0\in \mathbb R^8$
and $q_1 \in \mathbb C^8$ such that the following relations hold
\begin{equation} \label{eq:eigenvectors}
    A q_0 = 0, \quad A q_1 = i \omega q_1, \quad A \bar{q}_1 = - i \omega \bar{q}_1, 
    \quad A^T p_0 = 0, \quad A^T p_1 = -i \omega p_1, \quad A^T \bar{p}_1= i \omega \bar{p}_1 \,.
\end{equation}
Furthermore, letting $\braket{a,b} = a^\dagger b$, the eigenvectors can be normalized such that
\begin{equation} \label{eq:normalization}
    \left< p_0, q_0 \right> = \left< p_1, q_1 \right> = \left< \bar{p}_1, \bar{q}_1 \right> =1 
\end{equation}
while the other scalar products vanish, thus $\braket{p_1,q_0} = \braket{p_1,\bar{q}_1} = 0$. Specifically, we may choose
\begin{gather}
    q_0 =
    \begin{pmatrix*}[r]
        \num{-0.024962045785115454}\\
        \num{-0.10198878031459158}\\
        \num{0.1370967174936959}\\
        \num{-0.18189204279387905}\\
        \num{0.10357353240934195}\\
        \num{0.39482963902880625}\\
        \num{0.386794832337773}\\
        \num{-0.7879510117934457}
    \end{pmatrix*}, \hspace{20mm}
    q_1 = \begin{pmatrix*}[r]
        \num{ 0.02737513360232479 - 0.003733614980153608i} \\
        \num{ 0.09744013935416052 - 0.0020530876830862745i} \\
        \num{-0.13774705391103442 + 0.00807224514050541i} \\
        \num{ 0.13175352387314418 + 0.07592000338211646i} \\
        \num{-0.10458024483382158 + 0.009853089597229372i} \\
        \num{-0.34394317550947723 - 0.06262404784966186i} \\
        \num{-0.3981806627081541  - 0.03840975492368228i} \\
        \num{ 0.8088879258111088  }
    \end{pmatrix*},
\end{gather}
and
\begin{gather}
    p_0 =
        \begin{pmatrix*}[r]
        0\\
        \num{290.40443359353833}\\
        \num{-27.716162413627636}\\
        \num{222.0377118032486}\\
        \num{197.5427565762014}\\
        \num{246.85342933411596}\\
        \num{65.72788403238042}\\
        \num{86.9898946889358}
    \end{pmatrix*}, \hspace{20mm}
    p_1 = \begin{pmatrix*}[r]
        0\\
        \num{134.55856247086953 + 29.9173893182363i} \\
        \num{-12.377125600554889 - 31.10832072754824i} \\
        \num{102.46219419760145 + 19.8244601587464i} \\
        \num{96.22303233249862 + 34.468566651217735i} \\
        \num{114.25684023697899 + 21.677063687988955i} \\
        \num{31.128205370978275 - 8.470440147865185i} \\
        \num{41.95818021284129 - 2.6264860002350456i}
    \end{pmatrix*}.
\end{gather}
Note that the above normalization does not uniquely define the eigenvectors.
For example, we can scale $(p_0, q_0) \rightarrow (c p_0, \frac1c q_0)$ for $c\neq 0$ while leaving
\eqref{eq:eigenvectors} and \eqref{eq:normalization} invariant.

To write the equations below in a compact form it is
convenient to introduce the matrices $\Phi = \left( q_0\; q_1\; \bar q_1
\right)$ and $\Psi = \left( p_0\; p_1\; \bar p_1 \right )$. Due to
the normalization \eqref{eq:normalization}, we have the identity $\left< \Psi,
\Phi \right> = \Psi^\dagger \Phi = I_3$, where $I_n$ is the $n\times n$ identity matrix. Moreover $P_c = \Phi \Psi^\dagger$ is a projector on the center subspace: $P_c^2=P_c$. Right at $\alpha = 0$ there exists a unique manifold tangent to $q_0$, $q_1$, and $\bar q_1$. For non-zero $\alpha$ this manifold can be continued into a one-parameter family of three-dimensional invariant manifolds known as the center manifold. Any real vector $\vec{g}\in\mathbb R^8$ that belongs to the central manifold can be represented as
\begin{equation}
    \vec{g} = x q_0 + z q_1 + \bar z \bar q_1 + w(x,z,\bar{z},\alpha\textbf{}),
\end{equation}
where $x=\left< p_0, g \right>\in\mathbb R$,
$z=\left< p_1, g \right>\in\mathbb C$, and $\left< p_j, w \right> = 0$ for $j=0,1$. The time derivatives of $x$ and $z$ are given by
\begin{equation} \label{eq:reducedsystem_complex_form}
    \begin{cases} 
    \begin{aligned}
        \dot x(t) &{}= f^0(x,z,\bar z, \alpha), \\
        \dot z(t) &{}= 
        f^1(x,z,\bar z, \alpha),
    \end{aligned}  
    \end{cases} 
\end{equation}
where 
\begin{equation} \label{eq:gj}
f^j(x,z,\bar z, \alpha) = \bar p_j \beta\textbf{}\Big(x q_0 + z q_1 + \bar{z} \bar{q}_1 + w(x,z,\bar z, \alpha) , \alpha\Big)\,, \qquad j=0,1.
\end{equation}
Meanwhile, $w$ cannot be an arbitrary function of its arguments but must satisfy the differential equation
\begin{equation} \label{eq:wdot}
    \dot w = 
    \left( I_8 -  P_c \right)\beta\Big(x q_0 + z q_1 + \bar{z}\bar{q}_1 + w(x,z,\bar z, \alpha) , \alpha\Big)\,.
\end{equation}
Next we expand the mappings $f^0, f^1$ and $w$ as follows
\begin{align}
f^0(x,z,\bar z, \alpha) 
=& \sum f^0_{ijklm} x^i z^j \bar{z}^k    \alpha_1^l \alpha_2^m \nonumber  \\
= &\,
f^0_{00010} \alpha_1 + f^0_{00001} \alpha_2 + \frac12 f^0_{200} x^2 + \Re \left(f^0_{020} z^2 \right) + f^0_{011} z\bar z + 2 x \Re \left( f^0_{110} z \right)  \nonumber \\
   &+  f^0_{10010} x \alpha_1 + f^0_{10001} x \alpha_2 + \frac16 f^0_{300} x^3 + f^0_{111} x |z|^2 +  x^2 \Re \left( f^0_{210} z \right) + |z|^2 \Re \left( f^0_{021} z \right),
 \notag    \vspace{2mm}  \\
f^1(x,z,\bar z, \alpha) =& \sum g^1_{ijklm} x^i z^j \bar{z}^k   \alpha_1^l \alpha_2^m   \nonumber\\
=&\, i \omega z+ f^1_{00010} \alpha_1 + f^1_{00001} \alpha_2 + \frac12 f^1_{200} x^2 + \frac12 f^1_{020} z^2 + \frac12 f^1_{002} \bar z^2 + f^1_{011} |z|^2 + f^1_{110} x z  \nonumber \\
                &+    f^1_{101} x \bar z + f^{1}_{01010} z \alpha_1 + f^{1}_{01001} z \alpha_2 + \frac16 f^1_{300} x^3 +
                    f^1_{111} x |z|^2 + \frac12 f^1_{210} x^2 z + \frac12 f^1_{021}  z |z|^2,
\notag    \\
w(x,z,\bar z, \alpha) =& \sum w_{ijklm} x^i z^j \bar{z}^k  \alpha_1^l \alpha_2^m   \label{eq:w} \\
= &\, w_{00010} \alpha_1 + w_{00001} \alpha_2 + \frac12 w_{200} x^2 + 2 \Re\left(w_{110} xz \right) +  w_{011} z\bar z + \Re\left( w_{020} z^2 \right),\notag 
\end{align}
where $f^{0,1}_\mu \in \mathbb C$ and $w_\nu \in \mathbb C^8$,
with multi-indices $\mu$ and $\nu$ parametrizing the degrees of expansion in couplings and in parameters. We also used an abbreviated notation where $f^{0,1}_{ijk} = f^{0,1}_{ijk00}$. Since $f^0$ is a real-valued
function, we have the identities $f^0_{ijkmn} = \bar f^0_{ikjmn}$ and 
$f^0_{ijjmn} \in \mathbb R$. Since $w$ is also a real-valued function, the
same symmetry holds for the coefficients of $w$.

Comparing the equation \eqref{eq:gj} and \eqref{eq:Fexpansion} we can fix the coefficients in \eqref{eq:w}:
\begin{align*}
    f^j_{200} &{}= \bar{p}_j \cdot \beta_{2,0}(q_0, q_0),
        &f^j_{110} &{}= \bar{p}_j \cdot \beta_{2,0}(q_0, q_1),
        &f^j_{020} &{}= \bar{p}_j \cdot \beta_{2,0}(q_1, q_1), \\
    f^j_{002} &{}= \bar{p}_j \cdot \beta_{2,0}(\bar q_1, \bar q_1),
        &f^j_{101} &{}= \bar{p}\textbf{}_j \cdot \beta_{2,0}(q_0, \bar q_1),
\end{align*}
for $j=0,1$. And
\begin{equation}
    f^j_{00010} \alpha_1 + f^j_{00001} \alpha_2 = \bar{p}_j \cdot \beta_{0,1} \alpha\,, \qquad j=0,1,
\end{equation}
For the higher order parameter-dependent terms $f^j_{10010}$ and $f^j_{10001}$ could be computed in a similar way.
Then we decompose $z = u + iv$, with $u$ and $v$ real, in \eqref{eq:reducedsystem_complex_form} to obtain the three-dimensional real system given by
\begin{equation} \label{eq:reducedsystem}
\begin{aligned}
    \dot x ={}& p_0 \cdot \beta_{0,1}\alpha + \frac12 f^0_{200} x^2 + f^0_{011} (u^2+v^2) + 2 (\Re(f^0_{110}) u - \Im(f^0_{110})) v x + \Re(f^0_{020}) (u^2-v^2) - \\
                    & 2\Im(f^0_{020}) u v + \Re(f^0_{10010}) x \alpha_1 + \Re(f^0_{10001}) x \alpha_2 + \frac16 f^0_{300} x^3 + \Re(f^0_{210}) u x^2 - \Im(f^0_{210}) v x^2 + \\ 
                    &\Re(f^0_{021}) u (u^2+v^2) - \Im(f^0_{021}) v (u^2+v^2) + \Re(f^0_{111}) x (u^2+v^2), \\
    \dot u ={}&  -\omega v + \Re (p_1) \cdot \beta_{0,1}\alpha + \frac12 \Re(f^1_{200}) x^2 + \frac12 (\Re(f^1_{020}) +  \Re(f^1_{002})) (u^2-v^2) + \\
                    & (\Im(f^1_{002}) -  \Im(f^1_{020})) u v + (\Re(f^1_{110}) +  \Re(f^1_{101})) x u + (\Im(f^1_{101}) - \Im(f^1_{110})) v x + \\
                    & \Re(f^1_{011}) (u^2+v^2) + \Re(f^1_{01010}) u \alpha_1 - \Im(f^1_{01010}) v \alpha_1 + \Re(f^1_{01001}) u \alpha_2 - \Im(f^1_{01001}) v \alpha_2 + \\
                    & \frac16 \Re(f^1_{300}) x^3 + \frac12 \Re(f^1_{210}) x^2 u - \frac12 \Im(f^1_{210}) x^2 v + \frac12 \Re(f^1_{021}) u (u^2+v^2) - \\
                    & \frac12 \Im(f^1_{021}) v (u^2+v^2) + \Re(f^1_{111}) x (u^2+v^2), \\
    \dot v ={}&   \omega u +  \Im (p_1) \cdot \beta_{0,1}\alpha+ \frac12 \Im(f^1_{200}) x^2 + \frac12 (\Im(f^1_{020}) +  \Im(f^1_{002})) (u^2-v^2) + \\ 
                    & (\Re(f^1_{020}) -  \Re(f^1_{002})) u v + (\Im(f^1_{110}) +  \Im(f^1_{101})) x u + (\Re(f^1_{110}) - \Re(f^1_{101})) x v + \\
                    & \Im(f^1_{011}) (u^2+v^2) + \Im(f^1_{01010}) u \alpha_1 + \Re(f^1_{01010}) v \alpha_1 + \Im(f^1_{01001}) u \alpha_2 + \Re(f^1_{01001}) v \alpha_2 +  \\
                    & \frac16 \Im(f^1_{300}) x^3 + \frac12 \Im(f^1_{210}) x^2 u + \frac12 \Re(f^1_{210}) x^2 v + \frac12 \Im(f^1_{021}) u (u^2+v^2) + \\
                    & \frac12 \Re(f^1_{021}) v (u^2+v^2) + \Im(f^1_{111}) x (u^2+v^2)\,.
\end{aligned}
\end{equation}
The truncated normal form \eqref{eq:ZHnormform} exhibits a rotational symmetry in the $(u,v)$-plane. This symmetry is not a symmetry of the full system. Consequently, the approximation $\eqref{eq:ZHnormform}$ misrepresents essential qualitative properties of the RG flow around the bifurcation point. For this reason we retained cubic terms in the above expansions so that the symmetry is broken. Incidentally, the cubic terms are non-resonant, meaning that one could get rid of them by a suitable change of variables and thereby restore the $(u,v)$ rotation symmetry at cubic order.  But in doing so, one generates new symmetry-breaking terms at higher order. In principle, one could retain the symmetry at any desired order by iterating this procedure, but retaining the symmetry at all orders requires a singular change of variables.

\bibliographystyle{ieeetr}
\bibliography{biblio}
\end{document}